\documentclass[10pt]{article}
\usepackage{amssymb}
\usepackage{amsmath}
\usepackage{amsthm}
\usepackage{latexsym}
\usepackage[dvips]{epsfig}
\usepackage{mathrsfs}
\usepackage{eufrak}

\theoremstyle{plain}
\newtheorem{proposition}{Proposition}

\newtheorem{theorem}{Theorem}
\newtheorem{assumption}{Assumption}

\font\SYM=msbm10
\newcommand{\Real}{\mbox{\SYM R}}
\newcommand{\Complex}{\mbox{\SYM C}}

\font\tenscr=rsfs10 scaled1100
\font\sevenscr=rsfs7 
\font\fivescr=rsfs5 
\skewchar\tenscr='177
\skewchar\sevenscr='177
\skewchar\fivescr='177
\newfam\scrfam
\textfont\scrfam=\tenscr
\scriptfont\scrfam=\sevenscr
\scriptscriptfont\scrfam=\fivescr

\def\O{\mathcal{O}}

\newcommand{\TT}[3]{T_{#1 \phantom{#2} #3}^{\phantom{#1} #2}}
\newcommand{\updn}[3]{#1^{#2}_{\phantom{#2}#3}}
\newcommand{\dnup}[3]{#1_{#2}^{\phantom{#2}#3}}

\begin{document}


\title{\textbf{Asymptotic properties of the development of conformally flat data near spatial infinity}}

\author{Juan Antonio Valiente Kroon \thanks{E-mail address:
 {\tt j.a.valiente-kroon@qmul.ac.uk}} \\
School of Mathematical Sciences,\\ Queen Mary, University of London,\\
Mile End Road,\\
London E1 4NS,
\\United Kingdom.}

\maketitle

\begin{abstract}
  Certain aspects of the behaviour of the gravitational field near
  null and spatial infinity for the developments of asymptotically
  Euclidean, conformally flat initial data sets are analysed.  Ideas
  and results from two different approaches are combined: on the one
  hand the null infinity formalism related to the asymptotic
  characteristic initial value problem and on the other the regular
  Cauchy initial value problem at spatial infinity which uses
  Friedrich's representation of spatial infinity as a cylinder.  The
  decay of the Weyl tensor for the developments of the class of
  initial data under consideration is analysed under some existence
  and regularity assumptions for the asymptotic expansions obtained
  using the cylinder at spatial infinity. Conditions on the initial
  data to obtain developments satisfying the Peeling Behaviour are
  identified. Further, the decay of the asymptotic shear on null
  infinity is also examined as one approaches spatial infinity. This
  decay is related to the possibility of selecting the Poincar\'e
  group out of the BMS group in a canonical fashion. It is found that
  for the class of initial data under consideration, if the
  development peels, then the asymptotic shear goes to zero at spatial
  infinity.  Expansions of the Bondi mass are also examined. Finally,
  the Newman-Penrose constants of the spacetime are written in terms
  of initial data quantities and it is shown that the constants
  defined at future null infinity are equal to those at past null
  infinity.
\end{abstract}


\section{Introduction}
Penrose's definition of \emph{asymptotic simplicity} is an attempt to
provide a characterisation of isolated systems in General Relativity
\cite{Pen63,Pen65a,PenRin86}. It offers a framework on which diverse
notions of physical interest can be defined and handled in a precise
and rigorous manner. There is, however, a notorious lack of nontrivial
examples where the associated formalism ---see e.g.
\cite{Ger76,NewTod80,PenRin86,Win80}--- has been actually been used to
extract some physics. The obvious reason for this is the scarcity of
radiative exact solutions to the Einstein field equations other than
the boost-rotation symmetric ones \cite{BicSch89a}. A more conspicuous
consideration in this regard is the fact that radiative spacetimes
have to be constructed starting from an initial value problem where,
say, some Cauchy initial data are provided. Up to fairly recently,
there has not been a way of relating the properties of the initial
data to the radiative properties of the development.  For example, how
does an initial data set have to be so that the resulting spacetime
\emph{peels}? ---that is, the components of the Weyl tensor in an
adapted gauge having a distinctive decay in terms of powers of an
affine parameter of the generators of outgoing light cones
\cite{PenRin86,Ste91}. Another natural question on these lines would
be how does the ADM mass and the Bondi mass relate? Similarly, is
there any relation between the angular momentum defined at spatial
infinity and that defined at null infinity? Which classes of initial
data allow to select the Poincar\'{e} group out of the group of
asymptotic symmetries ---the BMS group--- in a canonical fashion?

Arguably, there is a shortage of general results about the evolution of
initial data sets for the Einstein field equations near spatial infinity.
This particular has made the idea of relating the structure of initial
data with radiative properties a daunting endeavour. However, 
work by Friedrich \cite{Fri98a,Fri04} on the \emph{regular initial
  value problem at spatial infinity} for time symmetric initial data
sets has provided a tool to address questions similar to the ones
raised in the first paragraph through a systematic approach. Friedrich
has introduced a representation of spatial infinity as a cylinder
---\emph{the cylinder at spatial infinity}, $\mathcal{I}$--- in stark
contrast to the usual representation as a point. The cylinder at
spatial infinity can be regarded as a limit set of incoming and
outgoing characteristics of the Einstein field equations, and as such
it happens to be a \emph{total characteristic}. As such, it allows to
transport information from the Cauchy surface to null infinity without
being contaminated by any sort of boundary conditions.

The construction which led to the cylinder at spatial infinity allows
an unfolding of the evolution process, which in turn can be analysed
in all detail and to the desired order. The latter can be thought of
as enabling the construction of a certain kind of asymptotic
expansions which are completely determined by the Cauchy data near
infinity. Friedrich's seminal work has been extended in several
directions, automatising, in some sense, the calculation of
the asymptotic expansions of the relevant field quantities
\cite{Val04a,Val04d} and to include more general classes of initial
data sets \cite{Val04e,Val05a}. The analysis described in these
references has exhibited the existence of a hierarchy of obstructions
to the smoothness of null infinity which goes beyond that arising from
a mere consideration of the linear aspects of the field equations as described
in e.g. \cite{Fri98a,Val03a}. The existence of the hierarchy of
obstructions has nurtured a \emph{rigidity conjecture} on the
smoothness of null infinity which, broadly speaking, states that the
only Cauchy initial data sets which have a development with a smooth null
infinity\footnote{Here and in the sequel, \emph{smooth} will always mean
  $C^\infty$.} are those which are stationary close to infinity.
Similarly, a given degree of differentiability at null infinity, say
$C^k$, would only be possible if the initial data is stationary to a
given order ---which depends on $k$. The kind of initial data
required can be constructed by means of the gluing techniques of
\cite{Cor00,CorSch06,ChrDel03}. Intuitively, the absence of
gravitational radiation near spatial infinity which the rigidity
conjecture asserts, suggests that the incoming radiation has to die
off in the infinite future, and at the same time the system cannot
have emitted gravitational waves for all times in the infinite past.

Now, the calculation of the asymptotic expansions which arise from the
construction of the cylinder at infinity is performed in a
gauge\footnote{In this article, a \emph{gauge choice} is understood as
  a certain choice of coordinates, vector or spinorial frame, and also
  a choice of representative in the conformal class of the physical
  spacetime.} which is adapted to a Cauchy initial value problem
---remarkably, the gauge also allows to read off the structure and
location of infinity directly from the initial data.  On the other
hand, the discussion of the gravitational field near null infinity is
usually done using a gauge which is based on geometric structures of
null hypersurfaces. Having two different gauges, hampers, to a certain
extent, extracting the physics of the system. It also difficults the
attempts to assess the relevance of the presence of obstructions to
the smoothness to null infinity in the asymptotic expansions. One of
the aspects of the analysis of the asymptotic behaviour of the
gravitational field via the definition of asymptotic simplicity and
related constructions is that it allows to rephrase the decay of the
fields in terms of differentiability at the conformal boundary. Thus,
it is natural to expect that the presence of obstructions to
smoothness in the asymptotic expansions may result in a modified
peeling behaviour ---this idea has already been explored in
\cite{Val04b}.

A first study of the transformation between the gauge used in
Friedrich's analysis of spatial infinity and the gauge generally used
to discuss null infinity has been given in \cite{FriKan00}. In the
present article their analysis will be extended to the developments of
a class of initial data sets which are not time symmetric. The
consideration of initial data sets with a non-vanishing second
fundamental form will allow us to examine some time asymmetric aspects
of the structure of the conformal boundary of asymptotically flat
spacetimes which up to now have remained \emph{terra incognita.} On
the other hand, the discussion shall be restricted to conformally flat
initial data sets.  This is done for conciseness and for the ease of
calculations and presentation. Previous analysis with conformally flat
initial data sets ---see e.g. \cite{Val04a,Val04e}--- have shown that
most of the crucial phenomena and structure observed in this setting
admit a counterpart when discussing non-conformally flat data sets
---like in \cite{Val04d,Val05a}. Furthermore, the detailed
understanding of conformally flat initial data sets is of relevance in
view of their use in the numerical simulation of black hole spacetimes
---see e.g.  \cite{HanHusBru06} for a recent discussion on this. A
cautionary note is, however, due: most of the conformally flat initial
sets considered in the numerical simulations are boosted. This is done
on physical grounds. Here, initial data sets with linear momentum
---i.e.  boosted--- will not be considered. The reason being that, as
it will be discussed in the sequel, they are more technically involved
and less smooth. The effects of this lower regularity are of interest
in themselves and will be discussed elsewhere.

The article is structured as follows: section
\ref{section:basic_setup} discusses some ideas of the basis set up and
fixes some notation; section \ref{section:F-gauge} describes some
aspects of the ``cylinder at spatial infinity''-formalism; section
\ref{section:constraints} considers relevant aspects of the class of
initial data sets to be used; section \ref{section:expansions} muses
over the asymptotic expansions that can be calculated using the
cylinder at spatial infinity; section \ref{section:NP-gauge} goes
briefly about the construction of a frame adapted to null infinity
---the NP gauge--- while section \ref{section:relating_gauges}
discusses how it is related to the gauge used in the expansions of
section \ref{section:expansions}. This article revisits the analysis
carried out in \cite{FriKan00} of ``Bondi type'' systems ---what here
is called the Newman-Penrose gauge--- and combines it with the
asymptotic expansions for the development of conformally flat initial
data sets which have been calculated in \cite{Val04e}. In section
\ref{section:peeling} analysis of the implications of these expansions
and the related obstructions to the smoothness of null infinity is
done ---under some existence and regularity assumptions.  In section
\ref{section:sigma} the machinery is used to calculate expansions of
the spin coefficient $\sigma$ on null infinity in the NP gauge. The
behaviour of this coefficient is related to various physical aspects
of the theory of isolated systems. The implications of these
expansions are analysed. It is shown that if the spacetime peels, then
the behaviour of $\sigma$ is such that it is always possible to single
out the Poincar\'e group out of the asymptotic symmetry group in a
canonical way. Expansions of the Bondi mass are also considered. In
section \ref{section:NP-constants} the formalism is used to write down
the Newman-Penrose constants of the developments of the class of
initial data under consideration in terms of initial data quantities.
These expressions allow to show that the constants at future null
infinity are equal to those at past null infinity.

There is much more to the aforementioned analysis than what is
presented in this article. In particular, most of the calculations
involved in the analysis have been carried out using \emph{ex profeso}
scripts in the computer algebra system {\tt Maple V}. A first subset
of the scripts constructs the initial data; another subset determines
the coefficients of the asymptotic expansions in terms of initial data
quantities; finally a third subset calculates the transformation
between gauges and allows to calculate the asymptotic expansions of NP
objects.  In order to keep the presentation at a reasonable length,
the explicit expressions of intermediate steps will be omitted, and more
emphasis will be put on  to the qualitative aspects of the results. Some
familiarity with the use of spinors ---in particular the space-spinor
formalism of \cite{Som80}--- will be assumed. Details of the diverse
other formalisms used ---the cylinder at spatial infinity, the
formalism of null infinity, the construction of gauges, and how to
relate them--- is kept to the minimum necessary to motivate the
analysis and to point out the many subtleties arising.  In any case,
the reader is referred to the various references cited for more
complete details. Finally, are some concluding remarks
and a brief appendix containing some useful spinorial expressions.

\section{Basic setup} \label{section:basic_setup} Let
$(\mathcal{M},\widetilde{g}_{\mu\nu})$ be a vacuum spacetime arising
as the development of asymptotically Euclidean Cauchy initial data
$(\widetilde{S},\widetilde{h}_{\alpha\beta},\widetilde{\chi}_{\alpha\beta})$.
Let $C^\mu_{\phantom{\mu}\nu\lambda\rho}$ be the conformally invariant
Weyl tensor of the metric $\tilde{g}_{\mu\nu}$.  Tilded quantities
will refer to quantities in the \emph{physical spacetime}, whereas
  untilded ones will denote, generically, quantities in an
  \emph{unphysical} ---i.e. conformally rescaled--- spacetime. Indices
  $\mu,\nu,\lambda,\ldots$ (second half of the Greek alphabet) are
  spacetime indices taking the values $0,\ldots,3$; the indices
  $\alpha,\beta,\gamma,\ldots$ are spatial ones with range $1,2,3$.
  The Latin indices $a,b,c,\ldots$ will be used in spatial expressions
  which are valid for a particular coordinate system (usually a
  Cartesian normal one). The indices $i,j,k,\ldots$ are spatial frame
  indices, while $A,B,C,\ldots$ will be spinorial indices taking the
  values $0,1$.

The initial data sets are taken to be maximal ---that is,
$\widetilde{\chi}=\widetilde{\chi}^\alpha_{\phantom{\alpha}\alpha}=0$,
and the metric to be $\widetilde{h}_{\alpha\beta}$ negative definite ---the
latter in order to ease the later use of spinors. For simplicity and
definiteness, it shall be assumed that the initial hypersurface contains
only one asymptotic end ---that is, a region which is diffeomorphic to
$\Real^3$ minus a ball.

The analysis shall be concentrated in the region of spacetime near
null and spatial infinity. In order to discuss the behaviour of the
development of the initial data sets in the asymptotic region use will
be made of the so-called conformal picture. Let $i$ denote the point
at infinity corresponding to the asymptotic end under consideration.
The point $i$ is obtained by conformally compactifying the
hypersurface $\mathcal{S}$ with a conformal factor $\Omega$ which can
be obtained by solving the Einstein constraint equations ---see
section (\ref{section:constraints}). Let $\mathcal{S}$ be the
resulting compact 3-dimensional manifold. The manifold is identified
in a standard way with $\mathcal{S}\setminus\{i\}$. Most of the 
discussion will be carried out in a sufficient small neighbourhood 
$\mathcal{B}_a(i)$ of radius $a$ centred on $i$ and on the resulting
associated domains of influence and dependence
$\mathcal{N}=J^+(\mathcal{B}_a(i))\cup J^-(\mathcal{B}_a(i))$. 

As it is customary, the discussion of the initial data will be carried
out in terms of the tensors
\begin{equation}
h_{\alpha\beta}=\Omega^2
\tilde{h}_{\alpha\beta}=\vartheta^{-4}\tilde{h}_{\alpha\beta}, \quad
\chi_{\alpha\beta}=\Omega
\chi_{\alpha\beta}=\vartheta^{-2}\tilde{\chi}_{\alpha\beta}, \quad 
\psi_{\alpha\beta}=\Omega^{-1}\tilde{\chi}_{\alpha\beta}=\vartheta^2
\tilde{\chi}_{\alpha\beta}
\end{equation}
on the conformally rescaled manifold
$\mathcal{S}$. In this work attention will be restricted to
conformally flat initial data sets and their developments. This
assumption will prove fruitful and will allow to infer features
of the development of more general types of initial data sets. Under
these assumptions coordinates $x^a$ can be chosen such that:
\begin{equation}
h=-\delta_{ab} \mbox{d}x^a\mbox{d}x^b, \mbox{ in } \mathcal{B}_a(i). \label{conformally_flat_metric}
\end{equation}
Unless
otherwise stated, $\{x^a\}$ denote some normal coordinates with origin
at $i$ based on a h-orthonormal frame $e_j$.

\section{The Friedrich-gauge} \label{section:F-gauge} 
In \cite{Fri98a}
a certain representation of the region of spacetime
$\mathcal{N}=J^+(\mathcal{B}_a(i))\cup J^-(\mathcal{B}_a(i))$ ---which
can be thought of as that close to null infinity,
$\mathscr{I}=\mathscr{I}^-\cup\mathscr{I}^+$, and spatial infinity,
$i^0$--- has been introduced. The standard representation of this
region of spacetime depicts $i^0$ as a point. In contrast, the
representation of \cite{Fri98a} depicts spatial infinity as a cylinder
---the cylinder at spatial infinity. The technical and practical
grounds for introducing this description have been discussed at length
in the seminal reference. The original construction in \cite{Fri98a}
was carried out for the class of time symmetric metrics with analytic
conformal metric $h_{ab}$. However, as seen in \cite{Val04e,Val05a},
the construction can be adapted to settings without a vanishing second
fundamental form.

Friedrich's construction makes use of a blow up of the infinity of the
initial hypersurface to $S^2$. The blow up of $i$ to $S^2$ requires
the introduction of a particular bundle of spin-frames over
$\mathcal{B}_a(i)$.  Let $\tau_{AA'}$ be the spinorial version of a
timelike vector $\tau^\mu$. Assume the normalisation
$\tau_{AA'}\tau^{AA'}=2$. Let $SU(\mathcal{B}_a(i))$ denote the fibre
bundle of spinorial dyads $\{\delta_A\}_{A=0,1}=\{o_A,\iota_A\}$ over
$\mathcal{B}_a(i)$ with projection $\pi$, such that $
\tau_{AA'}=o_{A}\bar{o}_{A'}+\iota_{A}\bar{\iota}_{A'}$. The fibres of
$SU(\mathcal{B}_a(i))$ are of type $SU(2,\Complex)=\{ \updn{t}{A}{B}\in
SL(2,\Complex) \;|\;
\tau_{AA'}\updn{t}{A}{B}\updn{\bar{t}}{A'}{B'}=\tau_{BB'}\}$. With this notation, define the set
\begin{equation}
\mathcal{C}_a =\big\{ \delta(\rho,\updn{t}{A}{B})\in SU(\mathcal{B}_{a}(i)) \;\big|\; |\rho|<a, \; \updn{t}{A}{B}\in SU(2,\Complex)\big\},
\end{equation}
where $\rho$ is an affine parameter of horizontal curves on
$SU(\mathcal{B}_a(i))$ which project to geodesics on
$\mathcal{B}_a(i)$ starting at $i$. The matrices
$\updn{t}{A}{B}\in SU(2,\Complex)$ can be regarded as ``angular
coordinates''. Crucial for the applications is that,
$\mathcal{I}^0=\pi^{-1}(i)=\{\rho=0\}$, the fibre over $i$, is such
that $\mathcal{I}^0\simeq SU(2,\Complex)$.  Thus, working on $C_a(i)$
provides a representation of $\mathcal{B}_a(i)$ where $i$ has been
``blown up'' to a sphere. Note, however, that while $\mathcal{B}_a(i)$
is 3-dimensional, $\mathcal{C}_a$ is 4-dimensional. This extra
dimension can be interpreted as giving rise to the spin-weight of the
diverse spinorial objects when lifted from $\mathcal{B}_a(i)$ to
$\mathcal{C}_a$. This issue will arise again in section
\ref{section:relating_gauges}. 

\bigskip 
The conformal structure of the spacetime $(\mathcal{M},g)$ is exploited by the
introduction of conformal Gaussian coordinates based on a certain type
of conformal invariants: \emph{conformal geodesics} ---see e.g.
\cite{Fri03a}. Accordingly, the spinor field $\tau_{AA'}$ will be
thought of as being tangent to timelike conformal geodesics which are
orthogonal to $\mathcal{B}_a(i)$. For a gauge based on conformal
Gaussian coordinates, there exists a canonical conformal factor,
$\Theta$ ---i.e. a choice of representative of the conformal class of
the metric $g_{\mu\nu}$. It is given by
\begin{equation}
\Theta=\kappa^{-1}\Omega\left(1-\frac{\kappa^2\tau^2}{\omega^2}\right), \quad \mbox{ with }  \omega=\frac{2\Omega}{\sqrt{|D_\alpha\Omega D^\alpha \Omega|}}=\rho+\mathcal{O}(\rho),\label{Theta}
\end{equation}
where $\Omega=\mathcal{O}(\rho^2)$ is the conformal factor of the
initial hypersurface ---usually obtained from solving the constraint
equations---, $\kappa=\kappa^\prime \rho$, with $\kappa^\prime$
smooth, $\kappa'(i)=1$, and $\tau$ an affine parameter of the
conformal geodesics. The function $\kappa$ expresses the remaining
conformal freedom in the construction. For the ease of future
calculations the choice
\begin{equation}
\kappa=\omega,
\end{equation}
will be made. As it will be seen, with this choice, the \emph{locus}
of null infinity corresponds to $\tau=\pm 1$. Other choices are
possible, and indeed, the calculations in \cite{FriKan00} use the
apparently simpler $\kappa=\rho$ which renders a more complicated
expression for the locus of null infinity.

\bigskip 
In order to complete the construction of a representation of
$\mathcal{N}=J^+(\mathcal{B}_a(i))\cup J^-(\mathcal{B}_a(i))$ ---the region of
spacetime close to null and spatial infinity--- consider the set
$\mathcal{C}_{a,\kappa}=\kappa^{1/2}\mathcal{C}_a$ of scaled spinor
dyads $\{\kappa^{1/2}o_A, \kappa^{1/2}\iota_A\}$. Define the bundle
manifold
\begin{equation}
\mathcal{M}_{a,\omega}=\big\{ (\tau,q) \big |  q\in
\mathcal{C}_{a,\omega}, -1 \leq \tau \leq 1 \big\},
\end{equation}
see figure 1. The conformal factor $\Theta$ given by (\ref{Theta}) can
be used to define the following relevant sets:
\begin{subequations}
\begin{eqnarray*}
&& \mathcal{I}=\big \{(\tau,q)\in \mathcal{M}_{a,\omega} \;\big|\; \rho(q)=0, \;|\tau|<1\big\}, \\
&& \mathcal{I}^\pm= \big \{ (\tau,q)\in \mathcal{M}_{a,\omega} \;\big |\; \rho(q)=0, \;\tau=\pm1\big \}, \\
&& \mathscr{I}^\pm=\big\{ (\tau,q)\in \mathcal{M}_{a,\omega} \;\big |  \; q\in \mathcal{C}_{a,\omega}\setminus \mathcal{I}^0, \; \tau=\pm 1 \big\}, 
\end{eqnarray*}
\end{subequations}
which will be referred to as, respectively, the \emph{cylinder at
  spatial infinity}, the \emph{critical sets} where null infinity
touches spatial infinity and \emph{future and past null infinity}.

\begin{figure}[t]
\centering
\includegraphics[width=.4\textwidth]{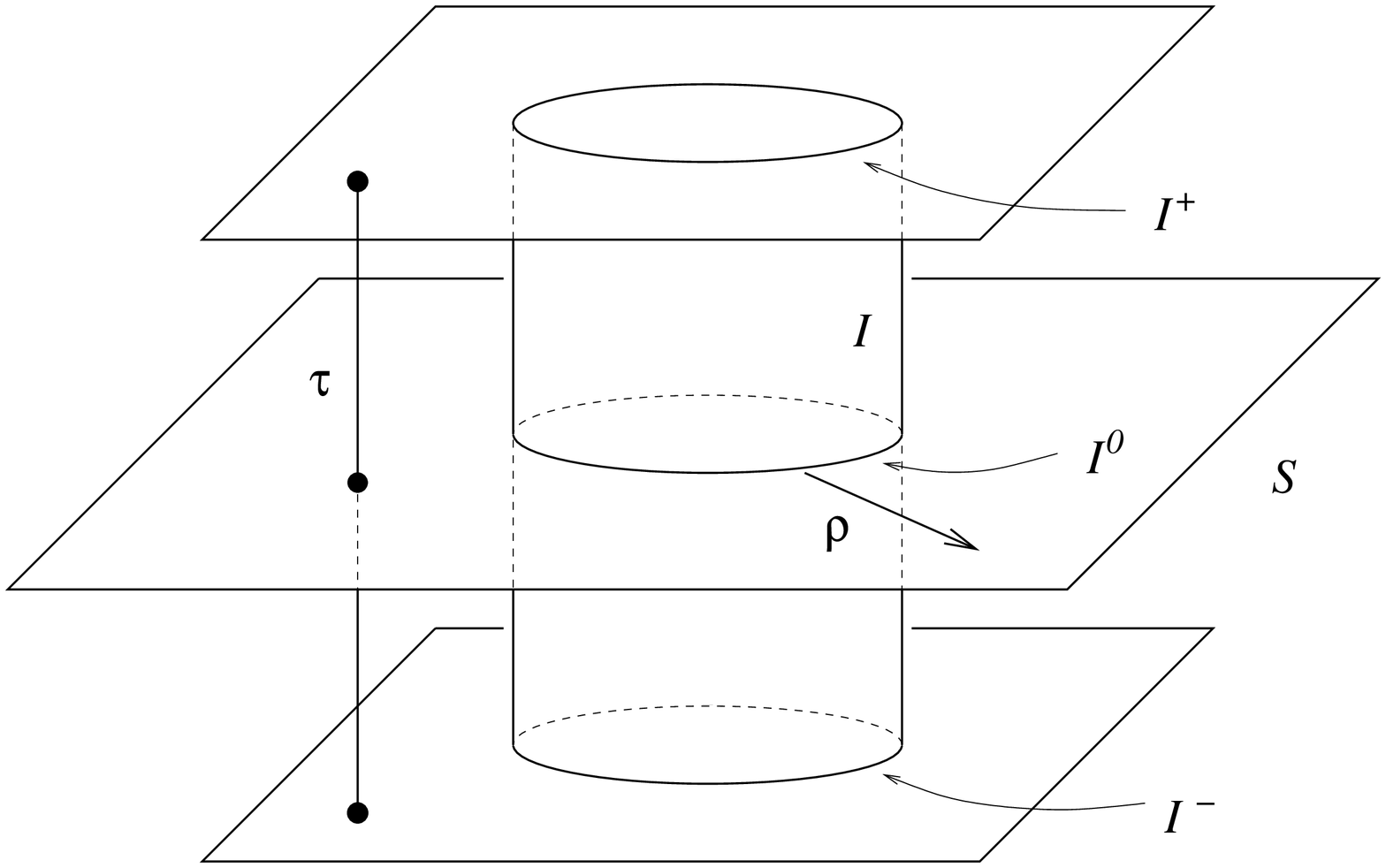}
\put(-20,100){$\mathscr{I}^+$}
\put(-25,15){$\mathscr{I}^-$}
\caption[]{}
\label{new:fig}
\caption{The region of the ---unphysical spacetime--- near spatial and null infinities. The conformal factor leading to this representation is given by $\Theta=\omega\Omega^{-2}(1-\tau^2)$.}
\end{figure}

Consistent with the structure of the bundle space
$\mathcal{M}_{a,\omega}$ it is possible to introduce a calculus based
on operators $X_+$, $X_-$ and $X$ ---related to a basis of the Lie
algebra $\mathfrak{su}(2,\Complex)$--- and the derivatives $\partial_\rho$,
$\partial_\tau$. For this, use will be made of a frame
$c_{AA'}$ and the associated spin connection coefficients
$\Gamma_{AA'BC}$ of $\nabla$. Conforming with the use of space-spinors
one writes
\begin{equation}
c_{AA'}=\tau_{AA'}\tau^{EE'}c_{EE'} -\tau^B_{\phantom{B}A'}c_{AB}, \quad c_{AB}=\tau_{(A}^{\phantom{(A}B'}c_{B)B'},
\end{equation}
with
\begin{equation}
\tau^{EE'}c_{EE'}=\frac{1}{\sqrt{2}}\partial_\tau, \quad c_{AB}=c^0_{AB}\partial_\tau + c^1_{AB}\partial_\rho+c^+_{AB}X_++c^-_{AB}X_-.
\end{equation}
It is also convenient to introduce an unprimed version of $\Gamma_{AA'BC}$,
\begin{equation}
\Gamma_{ABCD}=\dnup{\tau}{B}{B^\prime}\Gamma_{AB^\prime CD}, \quad \Gamma_{AA^\prime CD}=\Gamma_{ABCD}\updn{\tau}{B}{A^\prime},
\end{equation}
 which will be decomposed as
\begin{equation}
\Gamma_{ABCD}=\frac{1}{\sqrt{2}}(\xi_{ABCD}-\chi_{(AB)CD})-\frac{1}{2}\epsilon_{AB}f_{CD},
\end{equation}
where $\chi_{ABCD}$ agrees on $\mathcal{B}_a(i)$ with the second
fundamental form of the initial hypersurface.

The discussion of the curvature will make use of the spinor
$\Theta_{AA'BB'}$, related to the Ricci tensor of the Weyl connection
which can be constructed from $\nabla$ and a spinor $f_{AA'}$ with
spatial part given by $f_{AB}$ ---$f_{AA'}$ is related to the
acceleration of $\tau_{AA'}$. For later use, define
\begin{equation}
\Theta_{ABCD}=\dnup{\tau}{C}{A^\prime}\dnup{\tau}{D}{B^\prime}\Theta_{AA^\prime BB^\prime}.
\end{equation}
The remaining part of the curvature is encoded in the rescaled Weyl spinor $\phi_{ABCD}$ ---the spinorial counterpart of the rescaled Weyl tensor, $d^\mu_{\phantom{\mu}\nu\lambda\rho}=\Theta^{-1}C^\mu_{\phantom{\mu}\nu\lambda\rho}$. 

\begin{proposition} \label{proposition:F-gauge}
There is a frame $c_{AA'}$ on $\mathcal{M}_{a,\omega}$ for which 
\begin{subequations}
\begin{eqnarray}
&& \tau^{AA^\prime} c_{AA^\prime}=\sqrt{2}\partial_\tau, \\
&& \tau^{AA^\prime}f_{AA'}=0, \quad f_{AB}=-\tau^{CC^\prime}\Gamma_{CC^\prime AB}, \\
&& \tau^{BB^\prime}\Theta_{AA^\prime BB^\prime}=0, \quad \Theta_{ABC}^{\phantom{ABC}C}=0.
\end{eqnarray}
\end{subequations}
\end{proposition}

The description of $\mathcal{N}$ in terms of the
manifold $\mathcal{M}_{a,\omega}$, with the conformal factor $\Theta$
given by (\ref{Theta}), and the frame $c_{AA'}$ of proposition
\ref{proposition:F-gauge}, will be called the \emph{Friedrich gauge}
---F-gauge for short.

\bigskip In many parts of the sequel it will be necessary to
\emph{lift} objects defined in the base manifold $\mathcal{N}$ to the
manifold $\mathcal{M}_{a,\omega}$. Scalar and space spinorial objects
can be lifted in a natural way, with their ``angular dependence''
expressed in terms of the matrices $\updn{t}{A}{B}\in SU(2,\Complex)$.
Functions from $SU(2,\Complex)$ to $\Complex$ can be expanded in terms 
of certain functions $\TT{n}{l}{k}$ which can be thought of as the
lifts of spin-weighted spherical harmonics ---see \cite{Fri98a,Fri04}
for a more extended discussion:
\begin{equation}
{}_s Y_{nm}\mapsto (-\mbox{i})^{s+2n-m}\sqrt{\frac{2n+1}{4\pi}}\TT{2n}{n-m}{n-s}.
\end{equation}
The operators $X_+$, $X_-$ and $X$ can be defined by their action on
$\TT{n}{l}{k}$. The $\TT{n}{l}{k}$'s form a complete set of
orthonormal functions with respect to the Haar measure of
$SU(2,\Complex)$ ---this fact will be used several times.

\section{Conformally flat data without logarithmic singularities}
\label{section:constraints}

A discussion of conformally flat initial data which is well suited for
the purposes here presented has been given in \cite{DaiFri01}. Its
translation into the language of space-spinors has been carried out in
\cite{Val05a,Val06}. Here, the main results required are presented.

The space-spinorial versions of the momentum and Hamiltonian
constraints are given by
\begin{equation}
D^{AB}\psi_{ABCD}=0, \quad D^{AB} D_{AB}\vartheta =\frac{1}{8}\psi_{ABCD}\psi^{ABCD}\vartheta^{-7}.
\end{equation}
The spinorial translation of the approach adopted in \cite{DaiFri01}
---see e.g. \cite{Val04a}--- suggests a solution of the form ---see
the appendix for a definition of the irreducible spinors involved---:
\begin{eqnarray}
&&\psi_{ABCD}=\frac{\xi}{\rho^3}(3x_{AB}x_{CD}+h_{ABCD})+\frac{\bar{\eta}_1}{\rho^3}(x_{AB}y_{CD}+x_{CD}y_{AB})+\frac{\eta_1}{\rho^3}(x_{AB}z_{CD}+x_{CD}y_{AB}) \nonumber \\
&&\hspace{3cm}+2\frac{\bar{\mu}_2}{\rho^3}(y_{AB}y_{CD})+2\frac{\mu_2}{\rho^3} (z_{AB}z_{CD}), \label{psi_ABCD}
\end{eqnarray}
where
\begin{subequations}
\begin{eqnarray}
&& \xi= X_{-}^2 \lambda^R_2 +A +\rho Q,  \label{soln_lambda_1}\\
&& \eta_1=-2\rho\partial_\rho X_{-}\lambda^R_2 +X_{-}\lambda^I_2+\rho X_+Q+iX_+J, \label{soln_lambda_2} \\
&& \mu_2=2\rho\partial_\rho(\rho\partial_\rho \lambda_2^R)+X_+X_-\lambda_2^R-2\lambda_2^R-2\lambda^R_2-\rho\partial_\rho\lambda^I_2, \label{soln_lambda_3}
\end{eqnarray}
\end{subequations}
with\footnote{In older accounts ---see e.g. \cite{NewPen65}---, $\lambda^R_2$ and $\lambda^I_2$ are referred to as, respectively, the \emph{electric} and \emph{magnetic} parts of $\lambda_2$.}
\begin{equation}
\lambda_2=X_+^2\lambda=X_+^2\lambda^R+X_+^2\lambda^I=\lambda^R_2+\lambda^I_2,
\end{equation}
and $\lambda$ is the lift of a $C^\infty$ function on
$\mathcal{B}_a(i)\setminus \{i\}$. Note that an overbar denotes
complex conjugation. Finally,
\begin{subequations}
\begin{eqnarray}
&& J=3(\mbox{i}J_1-J_2)\TT{2}{0}{1}+3\mbox{i}J_3\TT{2}{1}{1}-3(\mbox{i}J_1+J_2)\TT{2}{2}{1}, \label{psi_J} \\
&& Q=\frac{3}{2}(\mbox{i}Q_2+Q_1)\TT{2}{0}{1}+\frac{3}{2}(Q_3)\TT{2}{1}{1}+\frac{3}{2}(\mbox{i}Q_2-Q_1)\TT{2}{2}{1}. \label{psi_Q}
\end{eqnarray}
\end{subequations}
In the above expressions, $A$, denotes a constant related to the
expansions conformal Killing vector. The term $J$ expresses the
angular momentum of the initial data, while $Q$ is associated with the
so-called conformal boosts. Note that boosted initial data sets
---i.e. with linear momentum--- are excluded from the class under
consideration. If the initial data is such that it has a non-vanishing
linear momentum, then it can be shown that the asymptotic expansion of
$\vartheta$ around infinity will contain logarithmic terms already at
order $\mathcal{O}(\rho^2)$. As firstly noted in \cite{Fri98a,Val04a},
even completely regular initial data sets ---e.g. time symmetric,
conformally flat--- may give rise to obstructions to the smoothness of
null infinity. It is expected that initial data with logarithms will
render further obstructions. This kind of obstructions arising from
particular non-smooth features of the data have not yet been studied.
Thus, attention is restricted to initial data sets which are as
regular as possible.

Even if the initial data sets are non-boosted, there is still
the possibility of having solutions to the Hamiltonian constraint
whose asymptotic expansions contain logarithms. The conditions under
which the solutions to the Euclidean momentum constraint give rise to
\emph{logarithm-free} solutions of the Hamiltonian constraint have
been spelled in \cite{DaiFri01}. From their results it follows that if 
\begin{equation}
\lambda=\frac{\lambda^\flat}{\rho}+\lambda^\natural, \label{lambda:Ansatz}
\end{equation}
where $\lambda^\flat$ and $\lambda^\natural$ are the lifts of
$C^\infty$ functions on $\mathcal{B}_a(i)$, then the asymptotic
expansion of the conformal factor $\vartheta$ contains no logarithms.
Recalling that the lift of an analytic function $f$ to $\mathcal{C}_a$
admits an expansion of the form
\begin{equation}
f=\sum_{p=0}^\infty \sum_{q\in Q(p)} \sum_{k=0}^{2q} f_{p;2q,k} \TT{2q}{k}{q}\rho^p,
\end{equation}
where $Q(p)=\{0,2,\ldots,p\}$ if $p$ is even and
$Q(p)=\{1,3,\ldots,p\}$ if $p$ is odd, the following assumption is made:

\begin{assumption} \label{assumption:initialdata} 
The class of initial
  data $(\mathcal{S},h_{\alpha\beta},\chi_{\alpha\beta})$ under
  consideration is conformally flat and with a solution to the
  momentum constraint given by (\ref{psi_ABCD}) where the lift of the
  complex function $\lambda$ to $\mathcal{C}_a$ is of the form
\begin{equation}
\lambda =\sum_{p=2}^\infty \sum_{q=2}^{p} \sum_{k=0}^{2q} \lambda_{p;2q,k} \TT{2q}{k}{q}\rho^{p}.
\end{equation}
\end{assumption}

For the class of initial data comprised by assumption
\ref{assumption:initialdata}, the lift of the conformal factor
$\vartheta$ to $\mathcal{M}_{a,\omega}$ admits the parametrisation
\begin{equation}
\vartheta=1/\rho+W,
\end{equation}
where 
\begin{equation}
W=\frac{m}{2}+\sum^\infty_{p=1}\frac{1}{p!}W_p, \quad W_p=\sum_{q=0}^p\sum_{k=0}^{2p}w_{p;2q,k}\TT{2q}{k}{q}, \label{W_definition}
\end{equation}
with $m$ the ADM mass, and $w_{p;2q,k}\in\Complex$ satisfying 
reality conditions so that $W=\overline{W}$.

\subsection{A further regularity condition}
\label{subsection:regularity_condition} 
Attention will be now given to a different type of regularity
condition on the initial data. In \cite{Fri98a} it has been shown
that, generically, a specific class of obstructions to the smoothness
of null infinity will be present unless certain coefficients in the
expansion of the spinor $\phi_{ABCD}$ on the initial
hypersurface, $\mathcal{S}$, satisfy a particular relation. In what
follows, objects expressible in terms of initial data which 
conrol the regularity of the development at null infinity will
be referred to as \emph{obstructions}. The obstructions described in
\cite{Fri98a} are related to the structure of the principal part of
the evolution equations (\ref{evolution_bianchi}), and are shared by
zero-rest-mass fields ---see \cite{Val03a}. For the case of time
symmetric data sets these conditions on the Weyl spinor can be
rewritten as a condition on the Cotton-Bach tensor. Namely:
\begin{equation}
D_{(A_pB_p}\cdots D_{A_1B_1}b_{ABCD)}(i)=0, \quad p=0,1,\ldots,
\end{equation}
where $b_{ABCD}$ is the spinorial version of the Cotton-Bach tensor.
The way the above formula has to be modified in order to accommodate
initial data sets which are not time symmetric is not known. However,
the conditions on the coefficients of the expansion of the Weyl tensor
given in \cite{Fri98a} can be calculated explicitly up to any desired
order for conformally flat initial data sets. One finds that ---cfr.
with \cite{Val05a}---:
\begin{equation}
\lambda_{p;2p,k}+\bar{\lambda}_{p;2p,k}=0, \label{lambda:conditions}
\end{equation}
with $p=2,3,\cdots$, $k=0,\ldots,2p$.

An explicit calculation shows that \emph{at least for $p=2,\ldots,6$}
the conditions (\ref{lambda:conditions}) are equivalent to
\begin{equation}
D_{(A_{p-1}B_{p-1}}\cdots D_{A_1B_1}C^\Re_{ABCD)}(i)=0, \quad p=1,\ldots,6, \label{regularity_C}
\end{equation}
 where
\begin{equation}
C^\Re_{ABCD}=\sqrt{2}D^E_{\phantom{E}(A}\chi^\Re_{BCD)E},
\end{equation}
$\chi^\Re_{ABCD}=\Omega^2\psi^\Re_{ABCD}$, with $\psi^\Re_{ABCD}$ being the part of the general solution to the Euclidean momentum constraint which is obtained arising from the real part of $\lambda$, that is, $\lambda^R$ \footnote{The tensorial equivalent of $C_{ABCD}$ is given by
\[
C_{kl}=D_i\chi_{j(k}\epsilon^{ij}_{\phantom{ij}l)}.
\]}. It is noted that for the class of data under consideration,
$C^\Re_{ABCD}(i)=0$, hence this particular condition poses no
restriction on $\lambda$. The equivalence of the conditions
(\ref{lambda:conditions}) and (\ref{regularity_C}) to all orders $p$
will be discussed elsewhere.

\section{Expansions near spatial infinity} \label{section:expansions}
The evolution of the spinorial quantities $c^\mu_{AB}$,
$\Gamma_{ABCD}$, $\Theta_{ABCD}$ and $\phi_{ABCD}$ describing the
geometry of the manifold $\mathcal{M}_{a,\omega}$ will be given by
means of the \emph{conformal propagation equations}
\cite{Fri98a,Fri04}. The procedure described in section
\ref{section:F-gauge}, by which the point $i^0$ is replaced by the
cylinder at spatial infinity $\mathcal{I}$, renders regular data and
equations for the unphysical spacetime. Moreover, it leads to an
unfolding of the evolution mechanism which allows to analyse the
process at arbitrary order and in all details. This kind of analysis
is made possible by the \emph{total characteristic} character of the
cylinder at spatial infinity with regard to the conformal propagation
equations ---$\mathcal{I}$ can be understood as a limit set of
incoming and outgoing light cones.

In order to briefly discuss the equations implied by the extended conformal equations it is convenient to write:
\begin{subequations}
\begin{eqnarray}
&& \upsilon=\big( c^\mu_{AB},\Gamma_{ABCD}, \Theta_{ABCD}\big), \\
&& \phi=(\phi_0,\phi_1,\phi_2,\phi_3,\phi_4),
\end{eqnarray}
\end{subequations} 
In this manner, one can write the propagation equations concisely in the form
\begin{subequations}
\begin{eqnarray}
&& \partial_\tau \upsilon =K\upsilon +G(\upsilon,\upsilon)+L\phi, \label{evolution_upsilon} \\
&& \sqrt{2}E\partial_\tau\phi+A^{AB}c^\mu_{AB}\partial_\mu=B(\Gamma_{ABCD})\phi, \label{evolution_bianchi}
\end{eqnarray}
\end{subequations}
where $K$, $G$ denote, respectively, linear and quadratic functions
with constant coefficients, $L$ denotes a linear function with
coefficients depending on the coordinates via $\Theta$ and related
objects, while $E$ denotes the $(5\times 5)$ unit matrix,
$A^{AB}c^\mu_{AB}$ are $(5\times 5)$ matrices depending on the
coordinates and $B(\Gamma_{ABCD})$ is a linear matrix-valued function
of the connection $\Gamma_{ABCD}$ with constant coefficients acting on
$\phi$. The reader is referred to \cite{Fri98a} for a full list of the
equations.

The evolution equations (\ref{evolution_upsilon}) and
(\ref{evolution_bianchi}) are to be supplemented with initial data
which can be readily calculated once an initial data set
$(\mathcal{S},h_{\alpha\beta},\chi_{\alpha\beta})$ satisfying the
vacuum constraint equations is known ---see \cite{Val04e} for the
details.

The unfolding of the evolution process referred to in the previous
paragraphs can be thought as the calculation of a certain type of
asymptotic expansions for the diverse field quantities grouped in the
vectors $\upsilon$ and $\phi$. Consistent with the class of
initial data under consideration ---which is expandable in terms of
powers of $\rho$--- one can put forward the following expansion Ansatz:
\begin{subequations}
\begin{eqnarray}
&& \upsilon \sim \sum_{p\geq 0} \frac{1}{p!}\upsilon^{(p)}\rho^p, \label{v_Ansatz}\\
&& \phi \sim \sum_{p\geq 0} \frac{1}{p!}\phi^{(p)} \rho^p. \label{phi_Ansatz}
\end{eqnarray}
\end{subequations}
The upper limits of the above summations have been intentionally left open to
emphasise some convergence issues which will be discussed later.
Transport equations for the above equations can be calculated by
differentiating the propagation equations
(\ref{evolution_upsilon})-(\ref{evolution_bianchi}) and then
evaluating at $\mathcal{I}$ ---that is, $\rho=0$. These equations are
schematically of the form
\begin{subequations}
\begin{eqnarray}
&&\partial_\tau v^{(p)} = Kv^{(p)}+G(v^{(0)},v^{(p)})+G(v^{(p)},v^{(0)}) \nonumber \\
&& \hspace{3cm}+\sum_{j=1}^{p-1}\left(G(v^{(j)},v^{(p-j)})+ L^{(j)}\phi^{(p-j)}\right) +
L^{(p)}\phi^{(0)}, \label{v_transport_eqns} \\
&&\big(\sqrt{2}E+A^{AB}(c^0_{AB})^{(0)}\big)\phi^{(p)} +
A^{AB}(c^C_{AB})^{(0)}\partial_C\phi^{(p)}= B(\Gamma^{(0)}_{ABCD})\phi^{(p)} \nonumber\\
&& \hspace{3cm} +\sum_{j=1}^p
\binom{p}{j}\left(B(\Gamma_{ABCD}^{(j)})\phi^{(p-j)}-A^{AB}(c^\mu_{AB})^{(j)}\partial_\mu
 \phi^{(p-j)}\right), \label{bianchi_propagation_transport_eqns}
\end{eqnarray}
\end{subequations}

These equations will be referred to as the \emph{transport equations
  of order} $p$. A crucial feature of the transport equations is that
their principal part is universal. The solution for $p=0$ is also
universal ---that is independent of the initial data--- and agrees
with the solution for Minkowski spacetime. For $p\geq 1$ they are
linear differential equations for the unknowns of order $p$.
Furthermore, note that the subsystem (\ref{v_transport_eqns}) consists
only of differential equations. The transport equations
(\ref{v_transport_eqns})-(\ref{bianchi_propagation_transport_eqns})
are decoupled in the following sense: knowledge of $\phi^{(j)}$,
$\phi^{(j)}$, $j=0,\ldots,p-1$, together with the initial data
$v^(p)|_{\mathcal{I}^0}$ allows us to solve the subsystem
(\ref{v_transport_eqns}) to obtain the quantities $v^{(p)}$. With
$v^{(k)}$, $\phi^{(l)}$, $k=0,\ldots,p$ and $l=0,\ldots,p-1$ and the
initial data $\phi^{(p)}|_{\mathcal{I}^0}$ at hand one could, in
principle, solve the equations
(\ref{bianchi_propagation_transport_eqns}) to get $\phi^{(p)}$.

In practise, in order to carry out the above described algorithm one
makes use of the fact that all the field quantities appearing in the
evolution equations
(\ref{evolution_upsilon})-(\ref{evolution_bianchi}) have a definite
spin-weight and expansion type. Thus, the coefficients $v^{(j)}$,
$\phi^{(l)}$, $0\leq j,l \leq p$ can be rewritten in terms of the
functions $\TT{n}{m}{k}$ and $\tau$-dependent coefficients. For
example, for the coefficients appearing in the components of the
rescaled Weyl tensor one writes:
\begin{equation}
\phi_j=\sum_{p\geq |2-j|} \sum_{q=|2-j|}^p \sum_{k=0}^{2q} \frac{1}{p!} a_{j,p;2q,k}(\tau) \TT{2q}{k}{q-2+j}\rho^p \label{phi_expansion}.
\end{equation}
The products of functions $\TT{n}{m}{k}$ which naturally arise in
(\ref{v_transport_eqns})-(\ref{bianchi_propagation_transport_eqns}) are
linearised by means of the use of the Clebsch-Gordan coefficients of
$SU(2,\Complex)$ ---note that the products appearing in the principal part of
the equations are somehow trivial as they involve the function
$\TT{0}{0}{0}$. The above ``decomposition in terms of spin-weighted
spherical harmonics'' allows for an implementation in a computer
algebra system\footnote{The computer algebra system {\tt maple V} has
  been used for this purpose.} whereby the diverse transport equations
in the hierarchy and their solutions are explicitly calculated to a
given order $p$. The order $p$ is restricted by the computer resources
available ---generally up to $p=7$ for a spacetime without axial symmetry.
The assumption of axial symmetry allows to carry the calculations up
to a couple of orders more.

An important question in the programme is the sense ---if any--- in which 
the expansions (\ref{v_Ansatz}) and (\ref{phi_Ansatz}) approximate the
solutions of the equations
(\ref{evolution_upsilon})-(\ref{evolution_bianchi}). In particular,
one would like to be able to estimate the residues
\begin{subequations}
\begin{eqnarray}
&& R_{p+1}(\upsilon)=\upsilon-\sum_{p^\prime=0}^p\frac{1}{p^\prime!}\upsilon^{(p^\prime)}\rho^{p^\prime}, \label{residue_v}\\
&&  R_{p+1}(\phi)=\phi-\sum_{p^\prime=0}^p\frac{1}{p^\prime!}\phi^{(p^\prime)}\rho^{p^\prime}. \label{residue_phi}
\end{eqnarray}
\end{subequations}
The estimation of the above residues and proving that the
asymptotic expansions actually do approximate the solution to the
conformal field equations is one of the remaining outstanding problems
in the analysis of the structure of spatial infinity. It is expected
that some suitable extension of the estimates discussed in
\cite{Fri03b} in the case of the spin-2 massless field (linearised
gravity) will permit to remove this hurdle. Some of the technical
complications of the problem lie in the degeneracy of the principal
part of the equations (\ref{evolution_bianchi}) and
(\ref{bianchi_propagation_transport_eqns}) at the critical sets
$\mathcal{I}^\pm$, which precludes the use of standard methods of the 
theory of partial differential equation. Further, as discussed in
\cite{Fri98a,Fri04,Val04a,Val04e,Val05a}, this degeneracy translates
in a non-smooth behaviour of the solutions of the transport equations
---which presumably will also extend to the actual solutions of the
field equations. It can be seen that for the data of assumption
\ref{assumption:initialdata} and if the further regularity conditions
(\ref{lambda:conditions}) hold, then up to $p=4$ the coefficients in
(\ref{v_Ansatz}) and (\ref{phi_Ansatz}) have all polynomial dependence
in $\tau$. Starting with $p=5$, logarithmic terms which are singular
at $\tau=\pm1$ appear. This adds further complexity to the analysis of
the asymptotic expansions. Hence, most of the calculations of
expansions are stopped once the first logarithmic divergences have
appeared.

In order to proceed with the analysis of this article, the following
assumption will be made:

\begin{assumption} \label{assumption:existence}
  For the class of initial data under consideration, the propagation
  equations (\ref{evolution_upsilon})-(\ref{evolution_bianchi}) have a
  unique solution in $\mathcal{M}_{a,\omega}$. Further, the expansions
  (\ref{v_Ansatz}) and (\ref{phi_Ansatz}) where their coefficients are
  given by the solutions of equations (\ref{v_transport_eqns}) and
  (\ref{bianchi_propagation_transport_eqns}) approximate these
  solutions in the sense that the residues (\ref{residue_v}) and
  (\ref{residue_phi}) can be suitably estimated and are finite in the
  region of interest. The regularity of the solutions is at least that
  of the $\tau$-dependence of the expansions.
\end{assumption}

\section{The Newman-Penrose gauge}
\label{section:NP-gauge}
This section discusses a certain class of gauge conditions on the
spacetime $(\mathcal{M},g_{\mu\nu})$ near each component of null
infinity ---i.e $\mathscr{I}^+$ and $\mathscr{I}^-$--- known as the
\emph{Newman-Penrose gauge} ---NP gauge. This class of gauge
conditions is convenient for the analysis of radiative properties of
the spacetime, like its peeling behaviour and its mass loss. The NP
gauge consists of certain requirements on the conformal gauge of the
spacetime, together with adapted coordinates and frame. The NP gauge
conditions have to be implemented independently for $\mathscr{I}^+$ and
$\mathscr{I}^-$. Here, for conciseness the discussion is concentrated on 
$\mathscr{I}^+$. The analogous construction for $\mathscr{I}^-$ can be obtained, in
principle, in the same way by interchanging the role of
the relevant vectors. The construction of the NP gauge is closely
related to the so-called \emph{asymptotic characteristic initial value
  problem} in which initial data is prescribed on an outgoing null
cone and a portion of $\mathscr{I}^+$ and allows to recover a portion of the
spacetime to the past of these null hypersurfaces ---see
\cite{Fri81,Fri81a,Fri81b,Fri82,Kan96b}. Alternatively, one could
prescribe data on an incoming null cone and on a portion of $\mathscr{I}^-$.
Here, however it will be assumed that the spacetime has been already
constructed, and a suitable conformal factor, $\Theta$, has been
found, allowing to locate the conformal boundary of the spacetime. 

Let $g_{\mu\nu}=\Theta^2 \tilde{g}_{\mu\nu}$ denote the metric of the
conformally rescaled (unphysical) spacetime. Our discussion will
consider frame fields $(e_{AA'})^\mu$ such that
$g(e_{AA'},e_{BB'})=\epsilon_{AB}\bar{\epsilon}_{A'B'}$. The
correspondence to the standard NP notation is given by
$(e_{00'})^\mu=l^\mu$, $(e_{11'})^\mu=n^\mu$, $(e_{01'})^\mu=m^\mu$
and $(e_{10'})^\mu=\bar{m}^\mu$. The spin connection coefficients
associated to the frame $(e_{AA'})^\mu$ are given by
\begin{equation}
\Gamma_{AA'BC}=\frac{1}{2}\big((e_{AA'})^\mu
(e_{B1'})^\nu
\nabla_\mu(e_{C0'})_\nu+(e_{AA'})^\mu
(e_{c1'})^\nu \nabla_\mu(e_{B0'})_\nu \big).
\end{equation}
The correspondence with the standard NP notation can is shown in the table 1.

\begin{table}
\label{table:spin_coefficients}
\begin{center}
\begin{tabular}{|c|c|c|c|}
\hline
                    & $BC=00$  & $BC=01$    & $BC=11$   \\
\hline
$AA^\prime=00^\prime$ & $\kappa$ & $\epsilon$ & $\pi$     \\
\hline
$AA^\prime=10^\prime$ & $\rho$   & $\alpha$   & $\lambda$ \\
\hline
$AA^\prime=01^\prime$ & $\sigma$ & $\beta$    & $\mu$     \\
\hline
$AA^\prime=11^\prime$ & $\tau$   & $\gamma$   & $\nu$     \\
\hline
\end{tabular}
\end{center}
\caption{The correspondence between the spin coefficients $\Gamma_{AA^\prime BC}$ and the standard NP notation.}
\end{table}
 
Consider a subset $\mathcal{N}^+\subset\mathcal{M}$ containing a
portion of $\mathscr{I}^+$. And let $\mathscr{C}^+$ be a \emph{fiduciary cut} of
null infinity: $\mathscr{C}^+\subset \mathscr{I}^+$, $\mathscr{C}^+\subset
\mathcal{N}^+$. It shall be assumed that $\mathscr{I}^+\cup\mathcal{N}^+\simeq
\mathscr{C}^+\times (-\varepsilon,\varepsilon)$, $\varepsilon>0$ ---that is,
it contains an open set of cuts. Note that because of the assumption
that the spacetime $\mathcal{M}$ arises as the development of some
asymptotically Euclidean initial data set, it follows that the cuts of
$\mathscr{I}^+$ have the topology of $S^2$.

\medskip
The following result shows that there is a certain gauge adapted to
the description of $\mathscr{I}^+$.

\begin{proposition} \label{proposition:np-gauge} 
Assume that
  $\mathcal{N}^+$ is suitably small. Then, there exists on
  $\mathcal{N}^+$ a conformal factor $\theta_+$, a frame
  $(e^+_{AA'})^\mu$ and coordinates $(u_+,r_+,x_+^2,x_+^3)$ such that
\begin{equation}
\Gamma^+_{10'00}=\bar{\Gamma}^+_{01'0'0'}, \quad
\Gamma^+_{11'00}=\bar{\Gamma}^+_{01'0'1'}+\Gamma^+_{01'01},
\quad \Gamma^+_{00'AB}=0, \quad \mbox{ on } \mathcal{N}^+,\label{np_conditions:N}
\end{equation}
and 
\begin{equation}
\Gamma^+_{10'11}=\Gamma^+_{11'11}=\Gamma^+_{01'11}=\Gamma^+_{11'01}=\Gamma^+_{11'00}=\Gamma^+_{10'00}=0 \quad \mbox{ on }\mathscr{I}^+\cap\mathcal{N}^+, \label{np_conditions:scri}
\end{equation}
where $\Gamma^+_{AA'BC}$ are the connection coefficients with
respect to the frame $(e^+_{AA'})^\mu$ of the Levi-Civita
connection $\nabla^+$ of the metric
$g^+_{\mu\nu}=(\Theta^+)^2\tilde{g}_{\mu\nu}$, where
$\Theta^+=\theta_+\Theta$. Furthermore,
\begin{equation}
R^+=0 \mbox{ on }\mathcal{N}^+
\end{equation}
and
\begin{equation}
\Phi^+_{22}=\frac{1}{2}\left(R^+_{\mu\nu}-\frac{1}{4}g^+_{\mu\nu}R^+\right)(e^+_{11'})^\mu (e^+_{11'})^\nu=0 \mbox{ on }\mathscr{I}^+\cap\mathcal{N}^+,
\end{equation}
where $R^+$ and $R^+_{\mu\nu}$ denote, respectively, the Ricci
scalar and Ricci tensor of the metric $g^+_{\mu\nu}$. The
coordinate $u_+$ on $\mathscr{I}^+$ is an affine parameter of the
generators of null infinity. The vector $(e^+_{11'})^\mu$ on
$\mathscr{I}^+$ is tangent to the generators of null infinity and satisfies
$e^+_{11'}(u_+)=1$. On $\mathcal{N}^+$, $u_+$
satisfies the eikonal equation $\nabla^{+\mu}u_+
\nabla^+_\mu u_+=0$, and the vector
$(e^+_{00'})^\mu$ is tangent to the generators of the
hypersurfaces $\mathcal{U}_{u_+}=\{ u^+=\mbox{constant}\}$
and $r^+$ an affine parameter. The coordinates $x_+^2$ and $x_+^3$
are chosen such that the metric induced on the cuts of $\mathscr{I}^+$ is
the standard $S^2$ metric.
\end{proposition}

\textbf{Remarks.}
\begin{enumerate}
\item Proofs of the above statements can be found in the
  discussions of \cite{FriKan00,Kan96b}.

\item The conformal factor, frame and coordinates given by the
  above result will be referred collectively to as the
  \emph{Newman-Penrose (NP) gauge for} $\mathscr{I}^+$.

\item A corresponding NP gauge consisting of a conformal factor
  $\theta_-$, a frame $(e^-_{AA'})^\mu$ and coordinates
  $(v_-,r_-,x_-^2,x_-^3)$ can be obtained by performing the
  interchange $0\mapsto 1$, $1\mapsto 0$ in the spinorial indices, so
  that e.g. $\Gamma^-_{01'00}=0$ on $\mathscr{I}^-\cap \mathcal{N}^-$ for
  some suitably small $\mathcal{N}^-\subset \mathcal{M}$. Here
  $\Gamma^-_{AA'BC}$ denote the spin coefficients with respect to
  $\{(e^-_{AA'})^\mu\}$ of the Levi-Civita connection $\nabla^-$ of
  the metric $g^-_{\mu\nu}=(\Theta^-)^2\tilde{g}_{\mu\nu}$, with
  $\Theta^-=\theta_-\Theta$.

\item The only non-vanishing coefficients on
  $\mathscr{I}^+\cap\mathcal{N}^+$ are $\Gamma^+_{01'00}$,
  $\Gamma^+_{01'01}$ and $\Gamma^+_{10'01}$. In particular
  $\sigma^+=\Gamma^+_{10'00}$ encodes information of physical
  interest.

\end{enumerate}

\section{Relating the F-gauge and the NP-gauge}
\label{section:relating_gauges} 
Having introduced two different gauges
for the analysis of the class of spacetimes under consideration, it is
now necessary to discuss the way they are related to each other.
The NP-gauge is hinged on null infinity and is adapted to
a characteristic initial value problem, while the F-gauge is based on
a Cauchy initial value problem. In the F-gauge the conformal factor
$\Theta$, and hence the location of null infinity is known \emph{a
  priori} by inspection of the initial data on $\mathcal{S}$ ---as
discussed in section (\ref{section:F-gauge}). Further, because of the
asymptotic expansions of section \ref{section:expansions} which allow
to examine the evolution of the field up to a given order $p$ in
terms of Cauchy initial data, the setting on which the F-gauge is
implemented will be regarded as the starting point.

As in the previous section, in order to fix ideas, the discussion will
be mainly concerned with the NP-gauge associated to $\mathscr{I}^+$.
A first important observation is that the construction of the F-gauge
and the associated regular Cauchy initial value problem near spatial
infinity ---see \cite{Fri98a,Val04a,Val04e,Val05a}--- lives in the
5-dimensional bundle manifold $\mathcal{M}_{a,\omega}$, and not in the
spacetime manifold, $\mathcal{M}$. The correspondence between objects on
$\mathcal{M}_{a,\omega}$ and $\mathcal{M}$ is carried out by the
projection $\pi$. Clearly, the correspondence between objects in the
two manifolds is not one to one. This ambiguity is related to the
spin-weight of the diverse spinorial objects under discussion ---put
in other way, when working on $\mathcal{M}_{a,\omega}$ one works with
all possible choices of frame vectors $(e_{01'})^\mu$,
$(e_{10'})^\mu$.

Assume that the region $\mathcal{N}^+$ considered in section
\ref{section:NP-gauge} is such that $\mathcal{N}^+\subset
\pi(\mathcal{M}_{a,\omega})$. In order to relate the frames
  $c_{AA'}$ ---living on $\mathcal{M}_{a,\omega}$--- and
  $e_{AA'}$ ---living on $\mathcal{N}^+$--- it is convenient to fix a
  section of $\mathcal{M}_{a,\omega}$ ---to be thought of as a choice
  of a tetrad $c^*_{AA'}$ on $\mathcal{N}^+$. Let $U$ be an open subset of $S^2$. Let $s:U\rightarrow SU(2)$ be a
local section of the Hopf fibration $SU(2)\rightarrow SU(2)/U(1)$.
The map $s$ induces a smooth section $U\times\Real\times\Real\ni
(p,\tau,\rho)\mapsto (s(p),\tau,\rho)\in \mathcal{M}_{a,\omega}$. The image
of $s$ will be denoted by $\mathcal{M}^*_{a,\omega}$. The vector fields
tangent to $s(U)$ which have a projection identical to that of
$X_\pm$ are of the form $X_\pm+a_\pm X$ with some functions $a_\pm$
on $s(U)$ which satisfy $a_-=-\bar{a}_+$. Due to the commutation
relations obeyed by $X_\pm$ and $X$, the functions $a_\pm$ do not
vanish on open subsets of $s(U)$. Thus, the fields $c^*_{AA'}$ on
$\mathcal{M}^*_{a,\omega}$ which satisfy $\pi(c^*_{AA'})=\pi_*(c_{AA'})$
are given by
\begin{equation}
c^*_{AA'}=c_{AA'}+(a_+c^+_{AA'}+a_-c^-_{AA'})X.
\end{equation}
The spin connection coefficients associated to the frame $c^*_{AA'}$ defined on $\mathcal{M}^*_{a,\omega}$ are given by
\begin{equation}
\Gamma_{AA'BC}^*=\Gamma_{AA'BC}
+(a_+c^+_{AA'}+a_-c^-_{AA'})(\epsilon_{0B}\epsilon_{C0}
-\epsilon_{1B}\epsilon_{C1}).
\end{equation}
In what follows, objects defined on $\mathcal{N}^+$ will be lifted to
the section $\mathcal{M}^*_{a,\kappa}$. In order to avoid complicating
the notation unnecessarily, lifted objects will be designed with the
same symbols as their original version on $\mathcal{N}^+$. From the
context it should be clear where the diverse objects live.

Let 
\begin{equation}
\Gamma^\star_{AA'BC}=\frac{1}{\theta_+}\left(\Gamma^*_{AA'BC}+ \epsilon_{A(B}c^*_{C)A'}(\ln \theta_+)\right),
\end{equation}
the spin connection coefficients associated to the frame
$c^\star_{AA'}=\theta^{-1}_+c^*_{AA'}$. From the discussion in
\cite{FriKan00} it follows:

\begin{proposition}
There is a Lorentz transformation, $\updn{\Lambda}{A}{B}\in SL(2,\Complex)$ such that  
\begin{equation}
e^+_{AA'}=\updn{\Lambda}{B}{A}\updn{\bar{\Lambda}}{B'}{A'} c^\star_{BB'}=\frac{1}{\theta_+}\updn{\Lambda}{B}{A}\updn{\bar{\Lambda}}{B'}{A'} c^*_{BB'},
\end{equation}
where $e^+_{AA'}$ and $\theta_+$ are the frame and conformal factor of proposition \ref{proposition:np-gauge}.
\end{proposition}

\textbf{Remarks.}
\begin{enumerate}
\item The matrix elements $\updn{\Lambda}{A}{B}$ on $\mathscr{I}^+$ can be
  determined entirely by means of a construction intrinsic to null
  infinity. Furthermore, for the class of spacetimes under
  consideration
\begin{subequations}
\begin{eqnarray}
\updn{\Lambda}{0}{1}|_{\mathscr{I}^+}=\rho^{1/2}+\mathcal{O}(\rho^{3/2}),&& \updn{\Lambda}{1}{1}|_{\mathscr{I}^+}=\frac{1}{2}X_+W_1 + \mathcal{O}(\rho^{5/2}), \\
\updn{\Lambda}{0}{0}|_{\mathscr{I}^+}=-\frac{101}{10}X_-W_1\rho^{3/2}+\mathcal{O}(\rho^{5/2}), && \updn{\Lambda}{1}{0}|_{\mathscr{I}^+}=-\rho^{-1/2}+\mathcal{O}(\rho^{1/2}),
\end{eqnarray}
\end{subequations} 
where $W_1$ is the $\mathcal{O}(\rho)$ term in the expansion of $W$
---see equation (\ref{W_definition}).  Latter discussion of the
Newman-Penrose constants will require the knowledge of the quantities
$e^+_{00'}(\updn{\Lambda}{A}{B})$. One has that
\begin{equation}
e^+_{AA'}(\updn{\Lambda}{B}{C})=-\updn{\Lambda}{F}{A}\updn{\bar{\Lambda}}{F^\prime}{A^\prime}\updn{\Lambda}{H}{C}\Gamma_{FF'\phantom{B}H}^{\star\phantom{F'}B}+\updn{\Lambda}{B}{D}\Gamma_{AA'\phantom{D}C}^{+\phantom{A'}D}.
\end{equation}
Remarkably, the gauge conditions (\ref{np_conditions:scri}) enable a
calculation of $e^+_{00'}(\updn{\Lambda}{A}{B})$ completely intrinsic
to $\mathscr{I}^+$.

\item The determination of the values of $\theta_+$ and
  $\mbox{d}\theta_+$ on $\mathscr{I}^+$ ---which is equivalent to the
  knowledge of the quantities $e^+_{AA'}(\theta)$--- can also
  performed in a way which is intrinsic to $\mathscr{I}^+$.

\item Consistently with the discussion of the previous section there is a function, $f_+$, on $\mathscr{I}^+$ such that
\begin{equation}
(e^+_{11'})^\mu=f_+ \nabla^{+\mu}\Theta^+ \mbox{ on }\mathscr{I}^+.
\end{equation}

\item The retarded time $u_+$ on $\mathscr{I}^+$ has the expansion
\begin{equation}
u_+|_{\mathscr{I}^+}=\sqrt{2}\left(-\frac{1}{\rho} +4m\ln \rho + u_* +\left( \frac{195}{28}m^2 -3A +\frac{74}{5}W_1\right)\rho +\mathcal{O}(\rho^2) \right),
\end{equation}
where $u_*$ is an arbitrary function of the angular coordinates
associated to the supertranslational freedom. This expansion has
been calculated up to $\mathcal{O}(\rho^4)$ inclusive. The retarded time $v_-$ has a similar expansion on $\mathscr{I}^-$:
\begin{equation}
v_-|_{\mathscr{I}^-}=\sqrt{2}\left(\frac{1}{\rho} -4m\ln \rho + v_* -\left( \frac{195}{28}m^2 +3A +\frac{74}{5}W_1\right)\rho +\mathcal{O}(\rho^2) \right).
\end{equation}
\end{enumerate}

In the sequel ---and in particular for the evaluation of the NP
constants--- it will be natural to push the fiduciary cuts of null
infinity to the critical sets where null infinity touches spatial
infinity. Thus, the following assumption is made.

\begin{assumption} \label{assumption:limits} The fiduciary cuts
  $\mathscr{C}^+$ and $\mathscr{C}^-$ of, respectively,
  $\mathscr{I}^+$ and $\mathscr{I}^-$ can be made coincide with the
  critical sets $\mathcal{I}^+$ and $\mathcal{I}^-$.
\end{assumption}

\section{Peeling properties} \label{section:peeling} 
In view of the existence of obstructions to the smoothness of null infinity
of the developments of the class of initial data sets of assumption
\ref{assumption:initialdata} discussed in sections
(\ref{subsection:regularity_condition}) and (\ref{section:expansions})
it is natural to ask how do they affect the asymptotic behaviour near
null infinity. The translation between gauges discussed in section
\ref{section:relating_gauges} provides an adequate way of performing this
discussion.

The \emph{Peeling Behaviour} is a specific type of decay of the
components of the Weyl tensor with respect to a frame satisfying the
NP-gauge. The classical discussion of the peeling behaviour is
generally carried out in the \emph{physical spacetime} ---a manifold
without boundary.  Let $\tilde{g}_{\mu\nu}$ the \emph{physical metric}
obtained from $g^+_{\mu\nu}$ via
\begin{equation}
\tilde{g}_{\mu\nu}=\tilde{r}^2 g^+_{\mu\nu},
\end{equation}  
where $\tilde{r}=1/r^+$. Coordinates in
$(\tilde{\mathcal{M}},\tilde{g}_{\mu\nu})$ can be introduced by
considering $(\tilde{r},u_+,x^2_+,x^3_+)$. Recall that $r^+=0$
---i.e. $\tilde{r}\rightarrow\infty$--- gives the locus of $\mathscr{I}^+$. A frame, $\tilde{e}_{AA'}$, and a spinor dyad, $\{\tilde{o}_+^A,\tilde{\iota}_+^A\}$  on the physical spacetime can be introduced via
\begin{subequations}
\begin{eqnarray}
&& \tilde{e}^+_{AA'}=\tilde{r}^{A+A'-2}e^+_{AA'}, \\
&& \tilde{o}_+^A=\tilde{r}^{-1} o_+^A, \quad \tilde{\iota}_+^A=\iota^A_+. 
\end{eqnarray}
\end{subequations}
In the coordinates $(\tilde{r},u_+,x^2_+,x^3_+)$ one has that
$\tilde{e}^+_{00'}=\partial_{\tilde{r}}$.

\bigskip
The original Peeling Behaviour described in the classical literature
assumes sufficient regularity of the solutions of the Einstein field
equations at, say, future null infinity so that in the physical
spacetime the components of the Weyl tensor admit an expansion of the
form ---see e.g. \cite{BonBurMet62,Sac62c,Pen63,Pen65a}:
\begin{equation}
\tilde{\Psi}_0= \mathcal{O}(\tilde{r}^5), \quad
\tilde{\Psi}_1= \mathcal{O}(\tilde{r}^4), \quad
\tilde{\Psi}_2= \mathcal{O}(\tilde{r}^3), \quad
\tilde{\Psi}_3= \mathcal{O}(\tilde{r}^2), \quad
\tilde{\Psi}_4= \mathcal{O}(\tilde{r}), \label{peeling}
\end{equation}
where
$\tilde{\Psi}_0=\Psi_{ABCD}\tilde{o}_+^A\tilde{o}_+^B\tilde{o}_+^C\tilde{o}_+^D$,
$\tilde{\Psi}_1=\Psi_{ABCD}\tilde{\iota}_+^A\tilde{o}_+^B\tilde{o}_+^C\tilde{o}_+^D$,
etc. The decay conditions for the components $\tilde{\Psi}_n$ are not
independent. Assuming a particular decay for $\tilde{\Psi}_0$, it is
possible to deduce the decay of the other components. Note that
\emph{smoothness at null infinity} is a much stronger requirement than
that of \emph{peeling behaviour} for which only a finite degree of
differentiability is needed.

In order to proceed some observations of generic nature are made. It
is recalled that the rescaled Weyl spinor, $\phi_{ABCD}$, is related
to the (conformally invariant) Weyl spinor $\Psi_{ABCD}$ by
\begin{equation}
\Psi_{ABCD}=\Theta\phi_{ABCD}, \label{Psi_phi}
\end{equation}
so that if $\Psi_{ABCD}=\mathcal{O}(\Theta)$, then
$\phi_{ABCD}=\mathcal{O}(\Theta^0)$. Further, the components
$\Psi_{ABCD}$ are related to those in the physical manifold via:
\begin{equation}
\tilde{\Psi}_n= \Theta^{4-n}\Psi_n. \label{rescaling_Psi}
\end{equation}
Moreover, it is noted that
\begin{equation}
\Theta=\O(1/\tilde{r}), 
\end{equation}
close to $\mathscr{I}^+$. 

Using the above relations, and assuming that assumption
\ref{assumption:existence} holds, it is possible to read off more or
less what the Peeling Behaviour of the developments of the initial
data sets of assumption \ref{assumption:initialdata} should be. It is
nevertheless necessary to analyse whether there occur any
cancellations which may improve  the decay  of the Weyl tensor.

\subsection{Data with $D_{(A_{p-1}B_{p-1}}\cdots D_{A_1B_1} C_{ABCD)}(i)=0$ for $p=1,2,3,4$.}

The best behaved case that will be considered here is that of an
initial data set where the regularity condition (\ref{regularity_C}) holds
up to $p=4$ inclusive. In this case, the main result of \cite{Val04e}
shows that all the functions $a_{j,p;2q,k}(\tau)$ appearing in the
expansions (\ref{phi_expansion}) up to $p=5$ are polynomials in $\tau$
save for the functions $a_{j,5;4,k}(\tau)$, $j=0,1,2,3,4$, with
$k=0,\ldots,4$ which contain logarithmic terms. More precisely, one has:
\begin{equation}
a_{j,5;4,k}=\Upsilon_{5;4,k}\bigg( (1-\tau)^{7-j} P_j(\tau) \ln(1-\tau) + (1+\tau)^{3+j}P_{4-j}(\tau)\ln(1+\tau)\bigg) +Q(\tau), \label{case1}
\end{equation}
where $\Upsilon_{5;4,k}$ are certain coefficients (obstructions)
completely specifiable in terms of the initial data, $P_j(\tau)$
denotes a certain polynomial of degree $j$ in $\tau$ such that $P_j(\pm
1)\neq 0$, and $Q(\tau)$ is an arbitrary polynomial.

Now, it is noted that from formula (\ref{Theta}) for the conformal factor
$\Theta$ it follows that close to, say $\mathscr{I}^+$,
\begin{equation}
(1-\tau)=\frac{\omega}{(1+\tau)\Omega}\Theta.
\end{equation}
Further, $(1+\tau)^{-1}$ is analytic close to $\tau=1$. Hence, a
direct use of the relations (\ref{case1}), (\ref{Psi_phi}) and
(\ref{rescaling_Psi}), shows that for this class of developments:
\begin{equation}
\tilde{\Psi}_n=\mathcal{O}\left(\frac{1}{\tilde{r}^{5-n}}\right),
\end{equation}
for $n=0,\ldots,4$. The first non-smooth terms in the expansions of
$\tilde{\Psi}_n$ arise in $\tilde{\Psi}_0$ at order
$\mathcal{O}(1/\tilde{r}^8)$, where a term of the form
\begin{equation}
\Upsilon_{5;4,k}\tilde{r}^{-8}\ln\tilde{r}, \label{singular_terms}
\end{equation}
$k=0,\ldots,4$, 
is to be found. It is  mentioned by passing that in this case the obstructions
$\Upsilon_{5;4,k}$ correspond to the polyhomogeneous constants of
motion discussed in \cite{ChrMacSin95,Val98,Val99a}.

\subsection{Data with $D_{(A_{p-1}B_{p-1}}\cdots D_{A_1B_1} C_{ABCD)}(i)=0$ for $p\leq 3$.}

As discussed in \cite{Val04a,Val04e}, the obstructions
$\Upsilon_{5;4,k}$ are special in the sense that they are the first
truly non-linear obstructions ---that is, they arise as a product of
the specific non-linear features of the conformal Einstein equations.
In contrast, the regularity conditions (\ref{regularity_C}) arise from
features which are present already at the linear level. Nevertheless,
it is of interest to analyse how these kind of obstructions affect the
peeling behaviour of the components of the Weyl tensor.

As discussed in \cite{Val04e}, if $\lambda_{p;2p,k}+\bar{\lambda}_{p;2p,k}=0$, 
for a certain $p$, then the functions $a_{j,p;2p,k}(\tau)$ in the expansion
(\ref{phi_expansion}) contain logarithmic terms. More precisely,
\begin{equation}
a_{j,p;2p,k}(\tau)= \Lambda_{p;2p,k}\bigg( (1-\tau)^{p-2+j}(1+\tau)^{p+2-j}\ln(1-\tau)
+(1+\tau)^{p-2+j}(1-\tau)^{p+2-j}\ln(1+\tau)\bigg)+ Q(\tau),
\end{equation}
with the obstruction $\Lambda_{p;2p,k}$ such that $\Lambda_{p;2p,k}=0$
if and only if $\lambda_{p;2p,k}+\bar{\lambda}_{p;2p,k}=0$. As
discussed in section \ref{subsection:regularity_condition}, for
$p=1,\ldots,6$ one can explicitly show that it is equivalent to the
condition (\ref{regularity_C}). It is plausible that this kind of
logarithms will propagate ---due to the hyperbolic nature of the
propagation equations--- and may produce further non-smoothness at
higher order. This process is, however, not yet understood.

According to the above discussion, if for $p=4$ the regularity
condition (\ref{regularity_C}) does not hold, then the resulting
peeling spacetime will contain ---in addition to the logarithmic terms
of (\ref{singular_terms})--- the following singular terms:
\begin{equation}
\Lambda_{5;10,k}\,\tilde{r}^{-8}\ln\tilde{r}.
\end{equation}
Note that in contrast to $\Upsilon_{5;4,k}$, which is an object of
quadrupolar nature, $\Lambda_{5;10,k}$ is of $2^5$-polar nature.

The above argumentation can be easily extended to the case where the
obstructions $\Lambda_{p;2p,k}$ with $p=2,3$ are present. In these
cases, the component $\tilde{\Psi}_0$ of the physical Weyl tensor
contains singular terms of the form
\begin{equation}
\Lambda_{p;2p,k}\,\tilde{r}^{-3-p}\ln\tilde{r}.
\end{equation}
It can be checked explicitly that when performing the Lorentz
transformation implied by the matrix $\updn{\Lambda}{A}{B}\in
SL(2,\Complex)$ of section \ref{section:relating_gauges}, there is no
cancellation taking place which would improve the regularity of the
components of the Weyl tensor. Thus, in the case $p=2$ the singular
terms arise sufficiently low in the expansions to preclude the peeling
behaviour.

One obtains the following. 

\begin{proposition}
Under assumption \ref{assumption:existence}, the developments of initial data sets of assumption \ref{assumption:initialdata} with
\begin{equation}
D_{(A_1B_1}C_{ABCD)}(i)=0
\end{equation}
peel in the sense that 
\[
\tilde{\Psi}_0= \mathcal{O}(\tilde{r}^5), \quad
\tilde{\Psi}_1= \mathcal{O}(\tilde{r}^4), \quad
\tilde{\Psi}_2= \mathcal{O}(\tilde{r}^3), \quad
\tilde{\Psi}_3= \mathcal{O}(\tilde{r}^2), \quad
\tilde{\Psi}_4= \mathcal{O}(\tilde{r}).
\]
On the other hand, if
\begin{equation}
D_{(A_1B_1}C_{ABCD)}(i)\neq 0
\end{equation}
then
\begin{equation}
\tilde{\Psi}_0=\mathcal{O}(\tilde{r}^{-5} \ln \tilde{r}),
\end{equation}
and accordingly the development does not peel.
\end{proposition}

\textbf{Remarks.}

\begin{enumerate}
\item Note that the requirements on conformally flat initial data to
  have peeling are very mild. In particular initial data sets like the
  Brill-Lindquist \cite{BriLin63}, Misner \cite{Mis63} and Bowen-York
  \cite{BowYor80} will satisfy them. Thus, modulo assumption
  \ref{assumption:existence}, the developments of these initial data
  sets will peel. However, the Weyl tensor of the unphysical spacetime
  will not be smooth due to the presence of the terms in
  (\ref{singular_terms}) ---the obstruction $\Upsilon_{5;4,2}$ has
  been explicitly been calculated for this case in \cite{DaiVal02} and
  shown to be non-zero unless the data is Schwarzschildean. These
  properties have been used in \cite{Val04b} to show the
  non-existence of conformally flat slices in the Kerr spacetime.

\item On the other hand, as shown in \cite{CouTor72}, the Einstein
  field equations are compatible with characteristic initial data with
  a decay of up to $\tilde{\Psi}_0=\mathcal{O}(\ln
  \tilde{r}/\tilde{r}^3)$. The question is, which kind of Cauchy data
  would produce such type of behaviour? From the discussion in this
  work, it is to be expected that conformally flat initial data with
  linear momentum and/or for which the complex function $\lambda$ has
  a singular behaviour worse than that of (\ref{lambda:Ansatz}) will
  give an example of initial data which would render this kind of
  spacetimes.

\item The definition of the Newman-Penrose constants ---see section
  \ref{section:NP-constants}--- requires certain degree of smoothness of
  the Weyl tensor. From the discussion presented in this section, the
  required condition on the initial data is
\begin{equation}
D_{(A_3B_3}D_{A_2B_2}D_{A_1B_1} C_{ABCD)}(i)=0,
\end{equation} 
so that 
\begin{equation}
\tilde{\Psi}_0=\mathcal{O}(\tilde{r}^{-7} \ln \tilde{r}).
\end{equation}
For spacetimes whose initial data do not satisfy the above it is not hard to see that the obstructions $\Lambda_{2;4,k}$ and $\Lambda_{3;6,k}$ are related to the conserved constants and logarithmic NP constants of \cite{ChrMacSin95,Val98,Val99a}.

\end{enumerate}

\section{Expansion of $\sigma$ on null infinity} \label{section:sigma}
As another application, the expansions of the Newman-Penrose
coefficient $\sigma^+=\Gamma_{10'00}^+$ on $\mathscr{I}^+$ ---the
\emph{asymptotic shear}--- are discussed.  The relevance of this stems
from three ---in principle unrelated--- directions. Firstly, as
discussed in \cite{NewPen66}, if $(\sigma^+)^R$ ---the electric part
of $\sigma^+$--- satisfies
\begin{equation}
\lim_{u_+\rightarrow -\infty}(\sigma^+)^R|_{\mathscr{I}^+}=0,
\end{equation}
that is, the limit as $\rho\rightarrow 0$, then there is a canonical
way of selecting the Poincar\'e group out of the asymptotic symmetry
group ---the Bondi-Metzner-Sachs (BMS) group. Secondly,
$\dot{\sigma}^+|_{\mathscr{I}^+}$ has been called in older accounts
the \emph{news function}, and enters in the expression of the mass
loss formula. Here and in the sequel, the overdot denotes
differentiation with respect to $u_+$. Thirdly, as discussed in
\cite{Chr91,Fra92}, $\sigma^+|_{\mathscr{I}+}$ is related to the
memory of non-linear tail waves first discussed by Christodoulou
\cite{Chr91,Fra92}.

Direct computer algebra calculations with the asymptotic expansions of
section \ref{section:expansions} and the transformation between the
F-gauge and the NP-gauge render the following result.

\begin{theorem} \label{theorem:sigma}
For the class of initial data of assumption \ref{assumption:initialdata} and under the existence assumptions \ref{assumption:existence} and \ref{assumption:limits} one has that 
\begin{equation}
\sigma^+|_{\mathscr{I}^+}=\mathcal{O}(\rho^3),
\end{equation}
as $\rho\rightarrow 0$. 
\end{theorem}

\textbf{Remarks.}
\begin{enumerate}
\item On the light of this result it is natural to ask what kind of initial
data is required to obtain a development such that the decay condition
\begin{equation}
\sigma^+|_{\mathscr{I}^+}\not\rightarrow 0 \mbox{ as }\rho\rightarrow 0, \label{bad}
\end{equation}
occurs. Preliminary calculations suggest that boosted conformally flat
initial data would give rise to developments for which the decay
(\ref{bad}) occurs. Another alternative would be to move away from the
choice (\ref{lambda:Ansatz}) of the complex function $\lambda$ giving
rise to the higher multipoles of the second fundamental form. In any
case, it seems that in order to obtain the asymptotic decay
(\ref{bad}) one has to consider solutions to the momentum constraint
with logarithms in their asymptotic expansions. These issues will be
explored elsewhere.

\item Another natural question to ask is: what is to be expected if
  one moves away from the restriction to conformally flat data? It is
  to be presumed that as long as one restricts to a class of
  ``reasonably smooth'' conformal metrics, $h_{\alpha\beta}$, the
  conclusion of theorem \ref{theorem:sigma} will still hold. As
  an example of what one would call ``reasonably smooth'' consider metrics
  with the same smoothness as the conformal metric of stationary
  solutions ---that is, $h_{\alpha\beta}\in
  C^{2,\alpha}(\mathcal{B}_a(i))$, where $C^{k,\alpha}$ denotes the
  standard H\"older space, see e.g.  \cite{Dai01b}.

\item The above two observations and the results of section
  \ref{section:peeling} give credence to the conjecture that \emph{any}
  peeling spacetime will have $\sigma^+|_{\mathscr{I}^+}\rightarrow 0$ as
  $u_+\rightarrow -\infty$.

\item If the initial data set satisfies the regularity condition
(\ref{regularity_C}) to some orders, then it is possible to provide
further information about the expansions of $\sigma^+$ on null
infinity. In particular, for a spacetime such that the first
obstructions to null infinity appear at order $p=4$ in the components
of the Weyl tensor, one has that
\begin{eqnarray}
&& \sigma^+|_{\mathscr{I}^+}=-\frac{1}{4}\bigg( \frac{1}{12}X_-^2 W_2 -(X_-W_1)^2 +\frac{15}{36}(X_-J)^2-32X_-W_1 X_-J \nonumber \\
&& \hspace{4cm} +\frac{8}{3}\sqrt{6}m (\bar{\lambda}_2-\lambda_2)-4\sqrt{6}(\bar{\lambda}_3-\lambda_3) \bigg)\rho^3+ \mathcal{O}(\rho^4). \label{sigma:expansion}
\end{eqnarray}
\end{enumerate}

From the expansion (\ref{sigma:expansion}) it is possible to derive an
expansion for the Bondi mass of the development, $m_B$, close to the
critical set $\mathcal{I}^+$. In the NP-gauge one has that
\begin{equation}
m_{B}=-\frac{1}{2}\oint_{\mathscr{C}_{u_+}}\bigg((\phi_{ABCD}\updn{\Lambda}{A}{1}\updn{\Lambda}{B}{1}\updn{\Lambda}{C}{0}\updn{\Lambda}{D}{0})|_{\mathscr{I}^+}+\sigma^+|_{\mathscr{I}^+}\dot{\bar{\sigma}}^+|_{\mathscr{I}^+}\bigg) \mbox{d}S.
\end{equation}
The above expression can be lifted in a natural way to the manifold
$\mathcal{M}_{a,\omega}$, by noting that surface element $\mbox{d}S$
has to be replaced with $\mbox{d}S\mbox{d}\alpha$ where $\alpha$ is a
parameter along the fibres. The functions $\TT{m}{k}{l}$ form an orthonormal basis with respect to  the Haar measure on
$SU(2,\Complex)$ which is given by $\mbox{d}\mu=(1/4\pi^2)\mbox{d}S\mbox{d}\alpha$, thus only terms with $\TT{0}{0}{0}$ contribute to the integration. One finds that:

\begin{proposition}
  For the class of initial data of assumption
  \ref{assumption:initialdata} and under the existence assumptions
  \ref{assumption:existence} and \ref{assumption:limits} one has
  that
\begin{equation}
\lim_{\rho\rightarrow 0} m_B =m.
\end{equation}
That is, the Bondi mass agrees with the ADM mass on the critical set
$\mathcal{I}^+$.
\end{proposition}

\textbf{Remarks.}

\begin{enumerate}

\item Versions of the above result with different regularity assumptions can be found in the literature ---see e.g. \cite{AshMag79,Her98}. The present form has been anticipated in \cite{FriKan00}.

\item Using the expansions (\ref{sigma:expansion}) ---which are valid for developments for which the first obstructions to the smoothness of null infinity appear at order $p=4$--- one can obtain a generalisation of the expansion for the Bondi mass given in \cite{Val03b}:
\begin{subequations}
\begin{eqnarray}
&& m_B=m-\frac{3\pi^2}{560}\sum_{k=0}^4 |H_k|^2\rho^7+\mathcal{O}(\rho^8), \\
&& \phantom{m_B}=m+\frac{3\pi^2}{560}\sum_{k=0}^4 |H_k|^2\frac{1}{u^7}+\mathcal{O}(1/u^8), 
\end{eqnarray}
\end{subequations}
where 
\begin{subequations}
\begin{eqnarray}
&& H_0=(m w_{2;4,0} -2 \sqrt{6}w^2_{1;2,0})-\frac{15}{2}\sqrt{6}(J_2+\mbox{i}J_1)^2 \nonumber \\
&& \hspace{3cm} +16\sqrt{6}w_{1;2,0}(J_2-\mbox{i}J_1) \nonumber \\
&&\hspace{3cm}+{16}m(\lambda_{2;4,0}-\bar{\lambda}_{2;4,0}) -{24}(\lambda_{3;4,0}-\bar{\lambda}_{3;4,0}), \\
&& H_1=(m w_{2:4,1} -4\sqrt{3} w_{1;2,0}w_{1;2,1})-15\sqrt{3}(J_1J_3+\mbox{i}J_2J_3) \nonumber \\
&&\hspace{3cm}-{16}\sqrt{3}\big( \mbox{i}J_3 w_{1;2,0}-(J_2-\mbox{i}J_1)w_{1;2,1} \big) \nonumber\\
&&\hspace{3cm}+{16}m(\lambda_{2;4,1}-\bar{\lambda}_{2;4,1}) -{24}(\lambda_{3;4,1}-\bar{\lambda}_{3;4,1}), \\
&& H_2=(m w_{2;4,2} -4w_{1;2,1}^2-4w_{1;2,0}w_{1;2,2})+{15}(J_3^2-J^2_1-J^2_2) \nonumber \\
&&\hspace{3cm}-{16}\big(\mbox{i}J_1(w_{1;2,2}-w_{1;2,0})-J_2(w_{1;2,0}+w_{1;2,2})+2\mbox{i}J_3w_{1;2,1}\big) \nonumber \\
&&\hspace{3cm}+{16}m(\lambda_{2;4,2}-\bar{\lambda}_{2;4,2}) -{24}(\lambda_{3;4,2}-\bar{\lambda}_{3;4,2}), \\
&& H_3=(m w_{2;4,3} -4\sqrt{3}w_{1;2,1}w_{1;2,2})+15\sqrt{3}(J_1J_3-iJ_2J_3)+ \nonumber \\
&&\hspace{3cm}-{16}\sqrt{3}\big( \mbox{i}J_3 w_{1;2,2}-(J_2+\mbox{i}J_1)w_{1;2,1} \big) \nonumber \\
&&\hspace{3cm}+{16}m(\lambda_{2;4,3}-\bar{\lambda}_{2;4,3}) -{24}(\lambda_{3;4,3}-\bar{\lambda}_{3;4,3}), \\
&& H_4=(m w_{2;4,4} -2\sqrt{6}w^2_{1;2,2}) +\frac{15}{2}\sqrt{2}(J_2-\mbox{i}J_1)^2 \nonumber \\
&& \hspace{3cm} +16\sqrt{6}w_{1;2,2}(J_2+\mbox{i}J_1) \nonumber \\
&&\hspace{3cm}+{16}m(\lambda_{2;4,4}-\bar{\lambda}_{2;4,4}) -{24}(\lambda_{3;4,4}-\bar{\lambda}_{3;4,4}).
\end{eqnarray}
\end{subequations}

\end{enumerate}

In \cite{Fra92}, Frauendiener has identified the non-linear memory
effect of tail waves in terms of the difference, $\Delta
\sigma^+|_{\mathscr{I}^+}$, between the value of $\sigma^+|_{\mathscr{I}^+}$ at a
cut of $\mathscr{I}^+$ given by $\mathscr{C}_{u_+}=\{u_+=\mbox{constant}\}$ and
its value at some fiduciary cut $\mathscr{C}^+$. In the current context it
is natural to let the critical set $\mathcal{I}^+$ to be this cut. In
this light theorem \ref{theorem:sigma} states that there is no
``residual'' distortion contained in the initial data, and that, in
principle there is no restriction to the possible values of $\Delta
\sigma^+|_{\mathscr{I}^+}$ which can be attained along $\mathscr{I}^+$. On the
other hand, if $\sigma^+|_{\mathscr{I}^+}\not\rightarrow 0$ as
$u_+\rightarrow 0$, then there are values of the memory effect which are
not accessible for $\Delta\sigma^+|_{\mathscr{I}^+}$, as it satisfies the equation
\begin{equation}
X_-^2 \Delta\sigma^+|_{\mathscr{I}^+} =\int_{-\infty}^{u_+} \dot{\sigma}^+|_{\mathscr{I}^+} \dot{\bar{\sigma}}^+|_{\mathscr{I}^+} \mbox{d}u'.
\end{equation}

\section{The Newman-Penrose constants} \label{section:NP-constants}

The Newman-Penrose constants are a set of nontrivial complex
quantities of quadrupolar nature defined on cuts of $\mathscr{I}^+$ (five of
them) and $\mathscr{I}^-$ (another five) which are absolutely conserved
---see \cite{NewPen65,NewPen68}. Here, absolutely conserved means that
their value is independent of the choice of cut, and that they are
supertranslation invariant. Further, the two sets of quantities
transform adequately under Lorentz transformations ---i.e. they form
a representation of the Lorentz group.

The argumentation leading to the NP constants makes certain
assumptions on the regularity of the spacetime at null infinity. Let
$\tilde{\Psi}_0=\Psi_{ABCD}\tilde{o}_+^A\tilde{o}_+^B\tilde{o}_+^C\tilde{o}_+^D$.

\begin{proposition}[Newman-Penrose, 1965]
Assume that on $\mathcal{N}^+\setminus (\mathcal{N}^+\cap\mathscr{I}^+)$ the component $\tilde{\Psi}_0$ of the Weyl tensor has the decay
\begin{equation}
\tilde{\Psi}_0=\Psi_0^{(5)} \tilde{r}^{-5}+\Psi_0^{(6)}\tilde{r}^{-6}+\O(\tilde{r}^{-7}\ln\tilde{r}), 
\end{equation}
then the quantities (the NP constants for $\mathscr{I}^+$)
 \begin{equation}
G^+_m=\oint_{\mathscr{C}^+} {}_2 \bar{Y}_{2m}\tilde{\Psi}_0^{(6)} \mbox{d}S,
\end{equation}
with $m=-2,\ldots,2$ are supertranslation invariant and absolutely
conserved ---that is, $\dot{G}^+_{m}=0$.
\end{proposition}

\textbf{Remarks.}
\begin{enumerate}
\item Note that the spacetime is required to have a bit more regularity at null infinity than what is needed for the peeling behaviour.

\item An analogous result holds for $\mathscr{I}^-$ if on $\mathcal{N}^-\setminus (\mathcal{N}^-\cap\mathscr{I}^-)$ the component $\tilde{\Psi}_4=\Psi_{ABCD}\tilde{\iota}_+^A\tilde{\iota}_+^B\tilde{\iota}_+^C\tilde{\iota}_+^D$ has the decay 
\begin{equation}
\tilde{\Psi}_4=\Psi_4^{(5)} \tilde{r}^{-5}+\Psi_4^{(6)}\tilde{r}^{-6}+\O(\tilde{r}^{-7}\ln\tilde{r}), 
\end{equation}
where here $\tilde{r}=1/r_-$. Then the NP constants for $\mathscr{I}^-$ are given by
\begin{equation}
G^-_m=\oint_{\mathscr{C}^-} {}_2 Y_{2m} \tilde{\Psi}_4^{(6)} \mbox{d} S.
\end{equation}

\item Note that in principle there is no reason to expect a relation between
$G^+_m$ and $G^-_m$, unless the spacetime is time reflexion symmetric
---like in the case analysed in \cite{FriKan00}.  

\end{enumerate}

A clear cut physical interpretation of the NP constants has remained,
so far, elusive. From their evaluation in particular examples and from
comparison with linear theories, it follows that the content of the
constants is nontrivial \cite{DaiVal02,FriKan00,NewPen68,LazVal00}.
There has been some work attempting to relate the constants with the
radiation content of the spacetime at late times and the value
of the Weyl tensor at timelike infinity, $i^+$ ---see e.g.
\cite{NewPen68,FriSch87,Val00a}. Remarkably, it has been recently
shown by Wu \& Shang \cite{WuSha07} that the NP constants vanish for
algebraically special spacetimes. This was a long-standing conjecture
---see \cite{Kin69}. From here it follows that the constants vanish
for the Schwarzschild spacetime (which was known) and for Kerr ---i.e.
for the (presumably!) asymptotic states of spacetimes containing
black holes. This result, together with conjectures given in
\cite{Val04e,Val05a} support the idea that the NP constants
measure in some way the gravitational radiation content of the
spacetime at earlier and later times.

In \cite{FriKan00} it has been shown that it is possible to reexpress
the NP constants in terms of quantities defined on the unphysical
manifold, $\mathcal{M}$. For this, note that $\phi_0=\tilde{\Psi}_{ABCD}o_+^A
o_+^B o_+^C o_+^D$, so that one obtains
\begin{equation}
G^+_{n}= -\frac{1}{2\pi}\oint_{\mathscr{C}^+} {}_2 \bar{Y}_{2n} e^{+}_{00'}(\phi_0) \mbox{d}S.
\end{equation}
In order to perform the obvious lift to the bundle
space $\mathcal{M}^*_{a,\omega}$ recall the correspondence
\begin{equation}
{}_2Y_{2n}\mapsto(-\mbox{i})^{2-n}\sqrt{ \frac{5}{4\pi}}\TT{4}{2-n}{0}, \quad {}_{-2}Y_{2n}\mapsto(-\mbox{i})^{2-n}\sqrt{\frac{5}{4\pi}}\TT{4}{2-n}{4}.
\end{equation}
If $\mbox{d}\mu$ denotes the Haar measure on $SU(2,\Complex)$, then $G^+_m$ lifts to
\begin{equation}
G^+_{n}= -\frac{(-\mbox{i})^{2-n}}{2\pi}\sqrt{\frac{5}{4\pi}}\oint_{\mathscr{C}^+} \TT{4}{2-n}{4} e^{+}_{00'}(\phi_0) \mbox{d}\mu \mbox{d}\alpha,
\end{equation}
where $\alpha$ denotes a parameter on the fibres of the fibre bundle
$\mathcal{M}_{a,\omega}$. The second integration does not alter the
result as the integrand is independent of $\alpha$. For the NP
constants on $\mathscr{I}^-$ one obtains in a similar way that
\begin{equation}
G^-_{n}= -\frac{(-\mbox{i})^{2-n}}{2\pi}\sqrt{\frac{5}{4\pi}}\oint_{\mathscr{C}^-} \TT{4}{2-n}{0} e^{-}_{11'}(\phi_4) \mbox{d}\mu \mbox{d}\alpha.
\end{equation}
Expanding the integrands one finds that
\begin{eqnarray}
&& \hspace{-5mm} G^+_n= -\frac{(-\mbox{i})^{2-n}}{2\pi}\sqrt{\frac{5}{4\pi}}\oint_{\mathscr{C}^+} \TT{4}{2-n}{4}\frac{1}{\theta^4}\bigg( \updn{\Lambda}{B}{0}\updn{\Lambda}{C}{0}\updn{\Lambda}{D}{0}\updn{\Lambda}{E}{0}\big(\updn{\Lambda}{A}{0}\updn{\bar{\Lambda}}{A'}{0'} c^*_{AA'}(\phi_{BCDE})-3\phi_{BCDE}e^+_{00'}(\theta) \big) \nonumber \\
&& \hspace{4cm}+4\theta
\updn{\Lambda}{B}{0}\updn{\Lambda}{C}{0}\updn{\Lambda}{D}{0}e^+_{00'}(\updn{\Lambda}{E}{0})\phi_{BCDE}\bigg)
d\mu d\alpha. \label{monster_1}\\
&& \hspace{-5mm}G^-_n= -\frac{(-\mbox{i})^{2-n}}{2\pi}\sqrt{\frac{5}{4\pi}}\oint _{\mathscr{C}^-}\TT{4}{2-n}{0}\frac{1}{\theta^4}\bigg( \updn{\Lambda}{B}{1}\updn{\Lambda}{C}{1}\updn{\Lambda}{D}{1}\updn{\Lambda}{E}{1}\big(\updn{\Lambda}{A}{1}\updn{\bar{\Lambda}}{A'}{1'} c^*_{AA'}(\phi_{BCDE})-3\phi_{BCDE}e^-_{11'}(\theta) \big) \nonumber \\
&& \hspace{4cm}+4\theta
\updn{\Lambda}{B}{1}\updn{\Lambda}{C}{1}\updn{\Lambda}{D}{1}e^-_{11'}(\updn{\Lambda}{E}{1})\phi_{BCDE}\bigg)
d\mu d\alpha. \label{monster_2}
\end{eqnarray}

The strategy is now to calculate 
\begin{equation}
G^+_n|_{\mathcal{I}^+}=\lim_{\rho\rightarrow 0} G^+_n, \quad G^-_n|_{\mathcal{I}^-}=\lim_{\rho\rightarrow 0} G^-_n,
\end{equation}
in order to obtain an expression for the NP constants in terms of
initial data quantities. A direct inspection using the results of the
asymptotic expansions discussed in section \ref{section:expansions}
reveals that there are potentially conflictive terms of order
$\mathcal{O}(\rho^{-1})$. Remarkably enough these cancel out. A computer algebra calculation renders the following result.

\begin{theorem}\label{theorem:equality}
  For the class of initial data given by (\ref{assumption:initialdata}) and
  under assumption \ref{assumption:existence}, the Newman-Penrose
  constants at $\mathscr{I}^+$ are equal to those at $\mathscr{I}^-$.
\end{theorem}

More precisely, one has that
\begin{equation}
G_n|_{\mathcal{I}^+}=G_n|_{\mathcal{I}^-}=\frac{\mbox{i}^{2-n}}{2}\sqrt{15\pi}G_n^\prime,
\end{equation}
where
\begin{subequations}
\begin{eqnarray}
&& G^\prime_0=127(m w_{2;4,0} -2 \sqrt{6}w^2_{1;2,0})-\frac{113}{2}(J_2+\mbox{i}J_1)^2 \nonumber \\
&& \hspace{3cm} +{240}w_{1;2,0}(J_2-\mbox{i}J_1) \nonumber \\
&&\hspace{3cm}+{120}m(\lambda_{2;4,0}-\bar{\lambda}_{2;4,0}) +{12}(\lambda_{3;4,0}-\bar{\lambda}_{3;4,0}), \\
&& G^\prime_1=127(m w_{2:4,1} -4\sqrt{3} w_{1;2,0}w_{1;2,1})-\frac{113}{2}\sqrt{12}(J_1J_3+\mbox{i}J_2J_3) \nonumber \\
&&\hspace{3cm}-{120}\sqrt{2}\big( \mbox{i}J_3 w_{1;2,0}-(J_2-\mbox{i}J_1)w_{1;2,1} \big) \nonumber\\
&&\hspace{3cm}+{120}m(\lambda_{2;4,1}-\bar{\lambda}_{2;4,1}) +{12}(\lambda_{3;4,1}-\bar{\lambda}_{3;4,1}), \\
&& G^\prime_2=127(m w_{2;4,2} -4w_{1;2,1}^2)+{113}(J_3^2-J^2_1-J^2_2) \nonumber \\
&&\hspace{3cm}-{240}\big(\mbox{i}J_1(w_{1;2,2}-w_{1;2,0})-J_2(w_{1;2,0}+w_{1;2,2})+2\mbox{i}J_3w_{1;2,1}\big) \nonumber \\
&&\hspace{3cm}+{120}m(\lambda_{2;4,2}-\bar{\lambda}_{2;4,2}) +{12}(\lambda_{3;4,2}-\bar{\lambda}_{3;4,2}), \\
&& G^\prime_3=127(m w_{2;4,3} -4\sqrt{3}w_{1;2,1}w_{1;2,2})+\frac{113}{2}\sqrt{12}(J_1J_3-iJ_2J_3)+ \nonumber \\
&&\hspace{3cm}-{120}\sqrt{2}\big( \mbox{i}J_3 w_{1;2,2}-(J_2+\mbox{i}J_1)w_{1;2,1}) \big) \nonumber \\
&&\hspace{3cm}+{120}m(\lambda_{2;4,3}-\bar{\lambda}_{2;4,3}) +{12}(\lambda_{3;4,3}-\bar{\lambda}_{3;4,3}), \\
&& G^\prime_4=127(m w_{2;4,4} -2\sqrt{6}w^2_{1;2,2}) +\frac{113}{2}(J_2-\mbox{i}J_1)^2 \nonumber \\
&& \hspace{3cm} +{240}w_{1;2,2}(J_2+\mbox{i}J_1) \nonumber \\
&&\hspace{3cm}+{120}m(\lambda_{2;4,4}-\bar{\lambda}_{2;4,4}) +{12}(\lambda_{3;4,4}-\bar{\lambda}_{3;4,4}).
\end{eqnarray}
\end{subequations}

It is noted that the integrand of the formula (\ref{monster_1}) can be written
concisely as:
\begin{eqnarray}
&& G=mW_2
+\sqrt{6}(X_-W_1)^2+\frac{113}{4572}\sqrt{6}(X_-J)^2-\frac{40}{127}\sqrt{6}X_-W_1
X_-J+\frac{120}{127}m(\lambda_2-\bar{\lambda}_2) \nonumber \\
&& \hspace{2cm} +\frac{12}{127}(\lambda_3-\bar{\lambda}_3).
\end{eqnarray}

From where it can more easily be seen that the constants have the structure:
\begin{eqnarray*}
&& (\mbox{mass})\times (\mbox{quadrupole})+(\mbox{dipole})^2+(\mbox{angular momentum})^2-(\mbox{dipole})\times (\mbox{angular momentum})\\
&& \hspace{4cm} +(\mbox{angular momentum quadrupole}).
\end{eqnarray*}

In the view of the above formulae, it is natural to ponder what would
happen if one were to consider a more general class of initial data
sets. It is to be expected that the theorem \ref{theorem:equality} will
still hold if one considers non-conformally flat initial data, for the
time asymmetry in the solutions of the Einstein constraint equations
is fed by the second fundamental form, not by the conformal metric. If
one were to consider boosted initial data, it is quite plausible that
null infinity may not be regular enough ---recall the existence of
logarithms in the initial data discussed in section
\ref{section:constraints}--- to ensure the constancy of $G^+_n$ and
$G^-_n$ ---unless, perhaps if the initial data set is not maximal,
like in the case of boosted slices of Schwarzschild.

\section{Concluding remarks}
The present article has been concerned with an analysis of the
physical implications of certain asymptotic expansions for
asymptotically flat spacetimes which allow to relate the behaviour
near null and spatial infinity to properties of the initial data.
These asymptotic expansions are obtained in a gauge which exploits to
the maximum the conformal structure of the spacetime. Hence, in order
to gain intuition on the physics of these expansions one has to
transform to a more ``physical gauge''. Crucial feature of the
expansions is that they bring to the foreground the fact that the
development of generic initial data will render a non-smooth conformal
boundary of the spacetime. This non-smoothness manifests itself in the
appearance of logarithmic terms in the solutions to the transport
equations which generate the terms in the asymptotic expansions. It
has to be emphasised that these logarithmic terms are conformal
invariants of the spacetime and not the product of some bad choice of
gauge. In other words one could completely forget about the F-gauge
and the cylinder spacetime, but one would still have to confront in
one way or another their implications ---as the discussion about the
peeling behaviour in section \ref{section:peeling} evidences. On a
more practical side, it is not clear what the implications for
numerical simulations of the non-smoothness of null infinity are. As seen
here, as long as no linear momentum is included in the reference
asymptotic end, the smoothness of most conformally flat initial data
sets of interest is enough peeling so that most of the $\mathscr{I}$
formalism remains untouched.

An interesting and in principle not evident result from our
calculations is the fact that the NP constants ---in the cases for which they
are well defined--- offer a set of absolute conserved quantities for
the whole spacetime, and not too different for each component of null
infinity. It has been argued that this feature will be preserved in
the case of more general initial data sets. Further, a simple
computation also reveals that the logarithmic NP constants should also
share this feature.

Finally, it has been seen that under certain existence assumptions
having the peeling behaviour automatically implies the right kind
of behaviour of the asymptotic shear at null infinity to obtain a
canonical representative of the Poincar\'e group out of the BMS group.
This result exemplifies the power and insights that can be gained by
methods presented in this work. At this point it would be natural to
proceed with a discussion of angular momentum. However, as the result
of the non-existence of conformally flat slices in the Kerr and other
(non-static) stationary spacetimes exposes, there is a certain 
non-cooperation between conformal flatness and the presence of angular
momentum at null infinity ---see e.g. \cite{Val04b,Val04c}. So it
seems that a more natural setting for this analysis is that of
non-conformally flat metrics.

This last issue and other questions raised in this article will be
further dwell upon elsewhere.

\section*{Acknowledgements}
I would like to thank H. Friedrich, S. Dain and R. Beig for valuable
discussions on diverse aspects of this research. I thank also CM
Losert-VK for a careful reading of the manuscript. I am
grateful to the Isaac Newton Institute, Cambridge for its hospitality
while some parts of this projects were in progress. This research is
funded by an EPSRC Advanced Research Fellowship.

\appendix

\section{Definitions of some spinors}
In the text, use of the following spinors has been made:
\begin{subequations}
\begin{eqnarray}
&& x_{AB}=\frac{1}{\sqrt{2}}(o_A\iota_B+\iota_A o_B), \quad y_{AB}=-\frac{1}{\sqrt{2}}\iota_A\iota_B, \quad z_{AB}=\frac{1}{\sqrt{2}}o_A o_B, \\
&& \epsilon^0_{ABCD}=o_A o_B o_C o_D, \quad \epsilon^1_{ABCD}= o_{(A} o_B o_C \iota_{D)}, \quad \cdots, \quad \epsilon^4_{ABCD}=\iota_A \iota_B \iota_C \iota_D, \\
&& h_{ABCD}=-\epsilon_{A(C}\epsilon_{D)B}.
\end{eqnarray}
\end{subequations}


\end{document}